\documentclass[useAMS,usenatbib,babel]{mn2e}

\usepackage[english,english]{babel}
\usepackage{amsmath}
\usepackage{amssymb,amsfonts,textcomp}
\usepackage{array}
\usepackage{supertabular}
\usepackage{hhline}
\usepackage{hyperref}
\usepackage[usenames]{color}
\hypersetup{dvips, colorlinks=true, linkcolor=black, citecolor=black, filecolor=black, urlcolor=black}
\usepackage[dvips]{graphicx}

\def\gtrsim{\lower.5ex\hbox{$\; \buildrel > \over \sim \;$}}
\usepackage{graphicx}

\newcommand{\mn}{\mbox{{\sc \small Horizon}-MareNostrum\,\,}}
\newcommand{\hagn}{\mbox{{\sc \small Horizon-AGN\,\,}}}

\definecolor{grey}{rgb}{0.75,0.75,0.75}
\definecolor{Orange}{rgb}{1.0,0.5,0.15}
\definecolor{brown}{rgb}{0.7,0.25,0.0}
\definecolor{pink}{rgb}{1.0,0.5,0.5}
\definecolor{darkerred}{rgb}{0.8,0,0}
\definecolor{darkerblue}{rgb}{0,0,0.8}
\definecolor{Blue}{rgb}{0,0.08,0.65}
\definecolor{Red}{rgb}{0.65,0.08,0.05}
\definecolor{Green}{rgb}{0.15,0.45,0.25}

\begin{document}

\author[Dubois, Y. et al. ]{
\parbox[t]{\textwidth}{Y. Dubois$^{1,2,3,4}$\thanks{E-mail: dubois@iap.fr},  C. Pichon$^{1,2,4,5}$, C. Welker$^{1,2,3}$, D. Le Borgne$^{1,2}$, J.~Devriendt$^{3,6}$, C.~Laigle$^{1,2}$,  S.~Codis$^{1,2}$, D.~Pogosyan$^{7}$, S.~Arnouts$^{8}$, K.~Benabed$^{1,2}$, E.~Bertin$^{1,2}$, J.~Blaizot$^{6}$,  F.~Bouchet$^{1,2}$, J.-F.~Cardoso$^{1,2}$, S.~Colombi$^{1,2}$, V. de Lapparent$^{1,2}$, V.~Desjacques$^{9}$, R.~Gavazzi$^{1,2}$,  S.~Kassin$^{10}$, T.~Kimm$^{11}$, H.~McCracken$^{1,2},$ B.~Milliard$^8$, S.~Peirani$^{1,2}$, S.~Prunet$^{1,2,12}$, S.~Rouberol$^{1,2}$, J.~Silk$^{1,2,3,13}$, A.~Slyz$^{3}$, T.~Sousbie$^{1,2}$, R.~Teyssier$^{14}$,  L.~Tresse$^8$, M.~Treyer$^8$, D.~Vibert$^8$ and M.~Volonteri$^{1,2,3,15}$}
\vspace*{6pt} \\
(Affiliations can be found after the references)
}
\date{Accepted 2014 June 19.  Received 2014 June 19; in original form 2014 January 29}

\title[Galactic spins within cosmic filaments]{
Dancing in the dark: galactic properties trace spin swings along the cosmic web}

\maketitle

\begin{abstract}
{A large-scale hydrodynamical cosmological simulation, \hagn, is used to investigate the alignment between the spin of galaxies and the cosmic filaments above redshift 1.2. The analysis of more than 150 000 galaxies per time step in the redshift range $1.2<z<1.8$ with morphological diversity shows that the spin of low-mass blue galaxies is preferentially aligned with their neighbouring filaments, while high-mass red galaxies tend to have a perpendicular spin. The reorientation of the spin of massive galaxies is provided by galaxy mergers, which are significant in their mass build-up. We find that the stellar mass transition from alignment to misalignment happens around $3\times 10^{10}\, \rm M_\odot$. Galaxies form in the vorticity-rich neighbourhood of filaments, and migrate towards the nodes of the cosmic web as they convert their orbital angular momentum into spin.  The signature of this process can be traced to the properties of galaxies, as measured relative to the cosmic web. We argue that a strong source of feedback such as active galactic nuclei is mandatory to quench \emph{in situ} star formation in massive galaxies and promote various morphologies. It allows mergers to play their key role by reducing post-merger gas inflows and, therefore, keeping spins misaligned with cosmic filaments.}
\end{abstract}


\section{Introduction}

Theoretical models of structure formation by gravitational instability
and numerical simulations have predicted that small fluctuations
from the early Universe
lead to the formation of a large-scale cosmic web made of clustered haloes,
filaments, sheets and voids~\citep[e.g.][]{zeldovichetal82,klypin&shandarin83, Blumenthaletal84, Davisetal85}.
The resulting properties of the Universe's large-scale structure
are the interplay of the planar  local collapse, as emphasized
in~\cite{zeldovich70} \citep[see also][]{shandarin&zeldovich89} and the inherent structure of the  Gaussian 
initial density and velocity shear fields,  leading to the cosmic web picture
of  dense peaks connected by filaments,  framing the  honeycomb-like structure 
of walls~\citep{bbks,bkp96}.

The extension of the Center for Astrophysics redshift survey~\citep{huchraetal83} gave spectacular observational
evidence~\citep{delapparentetal86,Gelleretal89} for this picture, 
triggering a renewed interest for such large-scale galaxy surveys~\citep{collessetal01, tegmarketal04}. 

Modern simulations have established a tight connection between the geometry and dynamics of the large-scale structure of matter, on the one hand, and the evolution of the physical properties of forming galaxies, on the other. 
Observational information on the morphology of galaxies and its dependence on environment is routinely becoming available for galaxies up to redshift two and beyond \citep{abrahametal07, oeschetal10, leeetal13}.
Matched samples at low and high redshifts allow for the study of the evolution of many physical properties of galaxies for most of the history of our Universe in unprecedented detail.
A key question formulated decades ago is nevertheless not satisfactorily answered: what properties of galaxies are driven by the cosmic environment?

There is ample literature~\citep[e.g.][]{hoyle49, peebles69, doroshkevich70, white84, schaefer09} on the tidal torque theory (TTT). 
It aims to explain the early acquisition of the spin\footnote{Hereafter, the spin is the angular momentum unit vector for simplicity.} of haloes, in the regime where the dynamics is well described by the Zel'dovich approximation, and when it is legitimate to assume that the tidal and the inertia tensors are uncorrelated.  
Within this framework, TTT predicts that the spin of haloes should be perpendicular to the direction of filaments.

Using \emph{N}-body simulations, which model only dark matter,~\cite{hahnetal07} and~\cite{zhangetal09} found halo spins preferentially oriented perpendicular to the filaments independent of halo mass.
 \citet{hatton&ninin01} claimed to detect an alignment between spin and filament, while \citet{faltenbacheretal02} measured a random orientation of the spins of haloes in the plane perpendicular to the filaments.
More recently a consensus  emerged when several works \citep{aubertetal04,bailin&steinmetz05, calvoetal07, hahnetal07b, pazetal08, sousbie08, libeskind12, trowlandetal13} reported that large-scale structures - filaments and sheets - influence the direction of the angular momentum of haloes in a way originally predicted by~\cite{sugermanetal00} and \cite{lee&pen00}. 
These studies pointed towards  a mass-dependent orientation of the spin, arguing for the first time that the spin of high-mass haloes tends to lie perpendicular to their host filament, whereas low-mass haloes have a spin preferentially aligned with it.
Nevertheless, the detected correlation remained weak and noisy until~\cite{codisetal12} confirmed it.
They quantified a redshift-dependent mass transition $M_{\rm tr, h}$, separating aligned from perpendicular haloes, and interpreted the origin of the transition in terms of large-scale cosmic flows.
\cite{codisetal12} found that high-mass haloes have their spins perpendicular to the filament because they are the results of mergers, a scenario suggested earlier by~\cite{aubertetal04} \citep[see also][]{bailin&steinmetz05}.
Low-mass haloes are not the products of mergers and acquire their mass by gas accretion in the vorticity-rich neighbourhood of filaments, which explain why their spins are initially parallel to the filaments~\citep{laigle2014, Libeskind13a}. 

\cite{Tempel13} recently found tentative evidence of such alignments in the Sloan Digital Sky Survey (SDSS) with an orthogonality for elliptical galaxies and a weak alignment for spiral galaxies~\citep[see also][]{tempel&libeskind13}. 
\cite{zhangetal13} found that the major axis of red galaxies is parallel to their host filaments and is the same for blue galaxies albeit with a weaker signature.
Similar measurements have been done for galaxies and walls; there is evidence that the spin of galaxies also lies within the walls in which they are contained~\citep{trujilloetal06}.

Besides those attempts to relate the spins of galaxies with the cosmic structure, much observational effort has been made to control the level of intrinsic alignments of galaxies as a potential source of systematic errors in weak gravitational lensing measurements~\citep[e.g.][]{C+M00,HRH00,H+S04}. 
Such alignments are believed to be the major source of systematics of the future generation of lensing surveys like \emph{Euclid} or Large Synoptic Survey Telescope (LSST).
Direct measurements of the alignment of the projected light distribution of galaxies in wide-field imaging data seem to agree on a contamination at a level between a few per cent and $\sim 10$ per cent of the shear correlation functions, although the amplitude of the effect depends on the population of galaxies considered~\citep[e.g.][]{L+P02,Joa++13a}. 
Given this dependence, it is difficult to use DM-only simulations as the sole resource to predict and control intrinsic alignments despite some success with the addition of a semi-analytical model prescription \citep[e.g.][]{Joa++13a}. 
The inherently anisotropic nature of the large-scale structure and its complex imprint on the shapes and spins of galaxies may prevent isotropic approaches from making accurate predictions.

Very few attempts have been made to probe the degree of correlation between \emph{galaxy} spins and their embedding cosmic web using hydrodynamical cosmological simulations.
\cite{hahn10} simulated the vicinity of a large-scale cosmic filament and found that -- at odds with the results presented in our paper -- the spin of galaxies within high-mass haloes is aligned with the filament while the spin of galaxies in low-mass haloes is perpendicular to the filament. 
\cite{gayetal10} focused on the colour gradients relative to the cosmic web using the \mn simulation~\citep{devriendtetal10}  which did not display much morphological diversity.
They found evidence of metallicity gradients towards and along the filaments and nodes of the cosmic web.
\cite{danovichetal11} also studied the feeding of massive galaxies at high redshift through cosmic streams using the \mn simulation. 
They found that galaxies are fed by one dominant stream (with a tendency to be fed by three major streams), streams tend to be co-planar (in the stream plane), and that there is a weak correlation between spin of the galaxy and spin of the stream plane at the virial radius, which suggests an angular momentum exchange at the interface between streams and galaxies (see also~\citealp{tillson12}).

In this paper, our focus is on the influence of the cosmic web as an anisotropic vector of the gas mass and angular momentum which ultimately shape galaxies.
Our purpose is to determine if the mass-dependent halo spin--filament correlations of~\cite{codisetal12} can be recovered via the morphology and physical properties of \emph{simulated galaxies}.
We aim to test these findings on a state-of-the-art hydrodynamical simulation, the so-called \hagn simulation, which produced over 150 000 resolved galaxies displaying morphological diversity by redshift $z=1.2$, in order to identify the effect of the environmentally driven spin acquisition on morphology and to probe the tendency of galaxies to align or misalign with the cosmic filaments as a function of galactic properties.

The paper is organized as follows. 
Section~\ref{section:virtual} describes the numerical set-up of our simulation, the post-processing of galaxy properties and filament tracing. 
Section~\ref{sec:result} presents the excess probability of alignment as a function of morphological tracers and investigates its redshift and spatial evolution. 
Section~\ref{sec:merger} discusses the origin of the observed (mis)alignment between galactic spins and filaments. 
We finally conclude in Section~\ref{sec:conclusion}.

\section{The virtual data set}
\label{section:virtual}

In this section, we describe the \hagn simulation (Section~\ref{section:numerics}), how galaxies are identified in it, and how the virtual observables are derived (Section~\ref{section:postprocess}). We conclude this section with a description of the tools used to compare the spins of galaxies to the orientation of the cosmic web (Section~\ref{section:skeleton}).

\subsection{The Horizon-AGN simulation}
\label{section:numerics}

\begin{figure*}
\includegraphics[width=17.5cm]{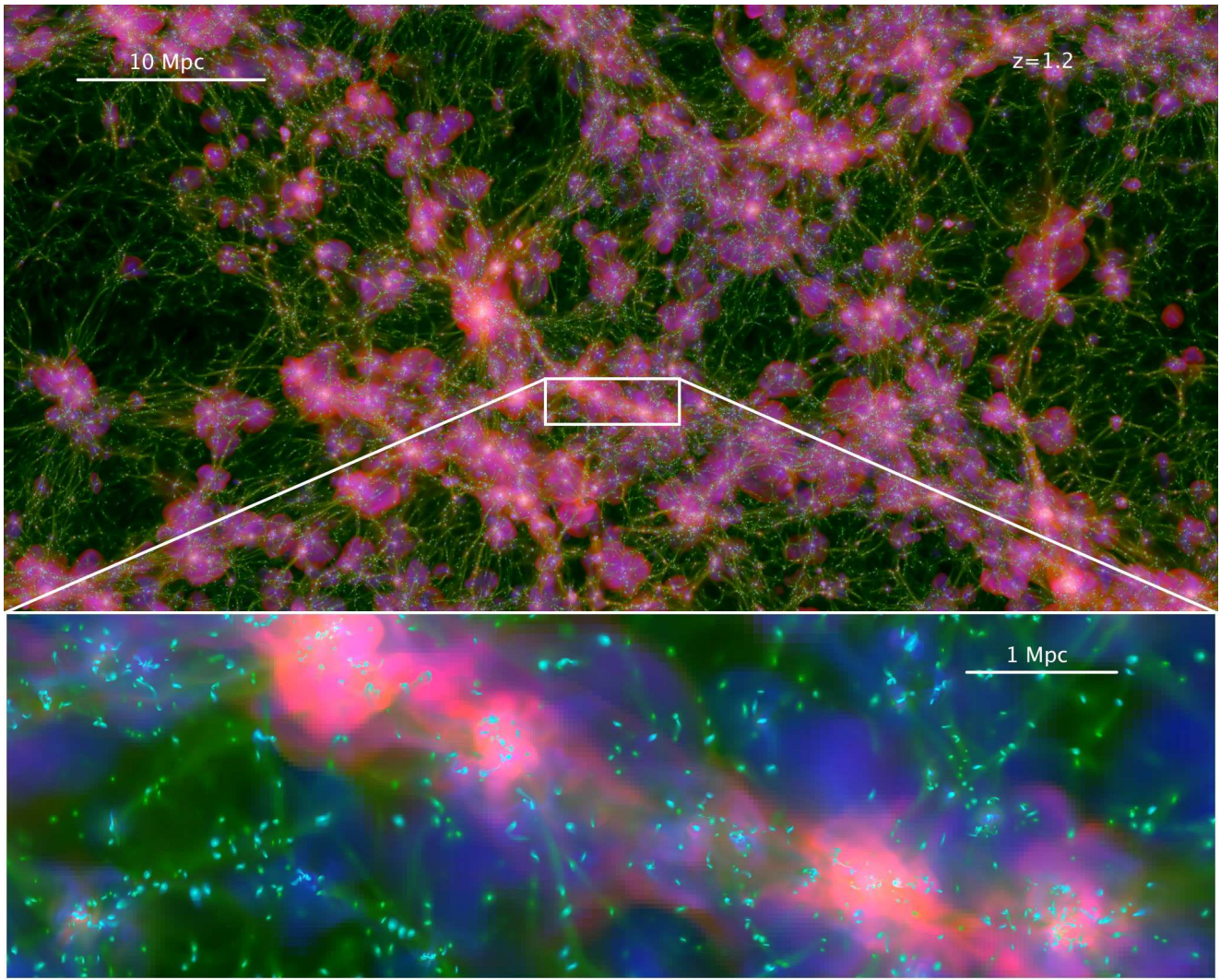}
        \caption{Projected maps of the \hagn simulation at $z=1.2$ are shown. Gas density (green), gas temperature (red) and gas metallicity (blue) are depicted. The top image is $100\,h^{-1}\, Ê\rm Mpc$ across in comoving distance and covers the whole horizontal extent of the simulation and $25\,h^{-1}\, Ê\rm Mpc$ comoving in depth. The bottom image is a sub-region where we see thin cosmic filaments as well as thicker filaments several Mpc long bridging shock-heated massive haloes and surrounded by a metal-enriched intergalactic medium. Physical scales are indicated on the figures in proper units.}
\label{fig:h-agn}
\end{figure*}

\begin{figure*}
\includegraphics[width=17.5cm]{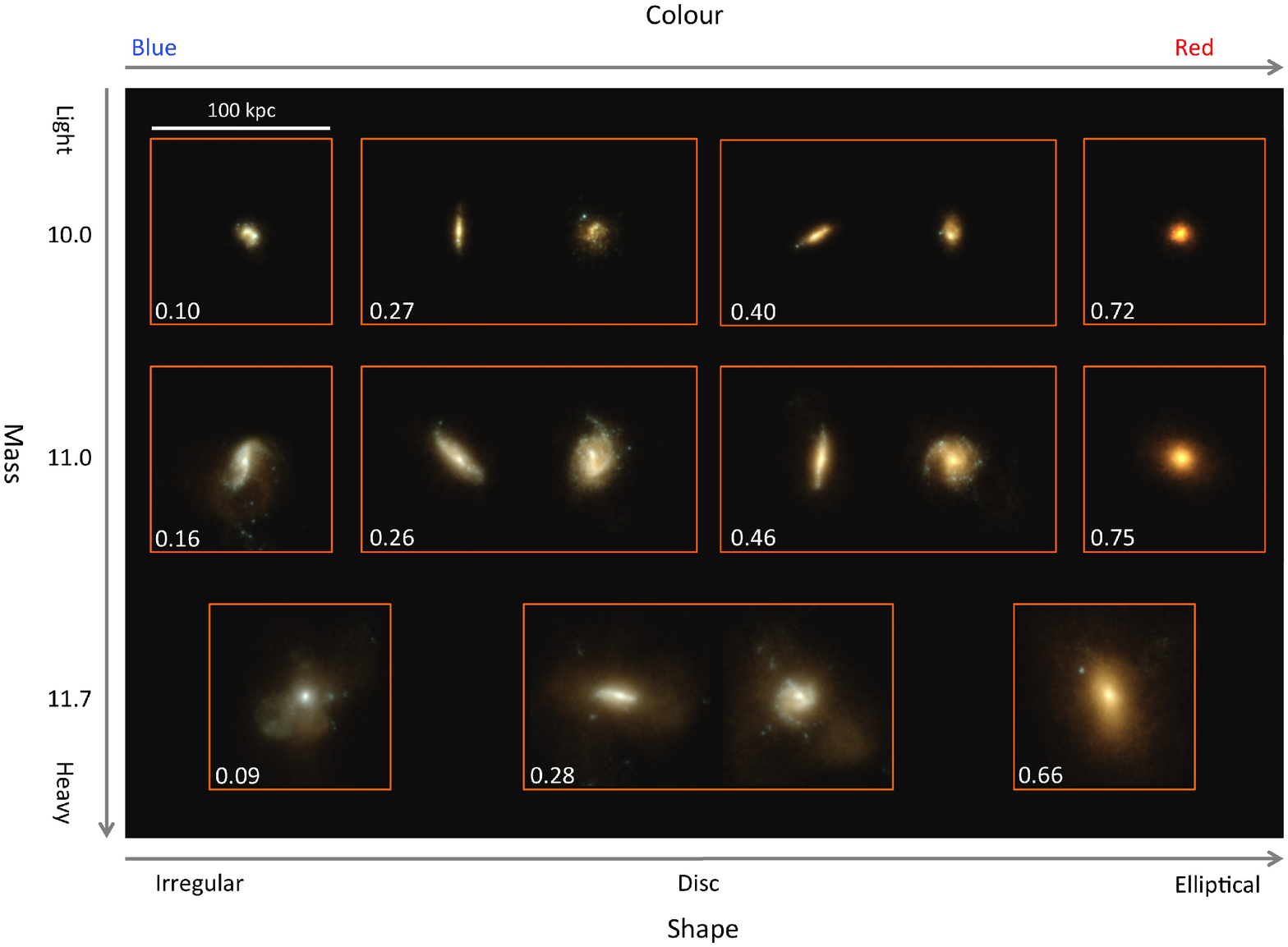}
\caption{Stellar emission of a sample of galaxies in the \hagn simulation at $z=1.3$ observed through rest-frame \emph{u}, \emph{g} and \emph{i} filters. Extinction by dust is not taken into account. Each vignette size is 100 kpc vertically. The numbers on the left of the figure indicate the galaxy stellar mass in log solar mass units. The number in the bottom left of each vignette is the \emph{g} -- \emph{r} rest-frame colour, not corrected for dust extinction. Disc galaxies (galaxies in the centre of the figure) are shown edge-on and face-on.}
\label{fig:h-agn_gal}
\end{figure*}

\subsubsection{The code and initial conditions}

We adopt a standard $\Lambda$ cold dark matter cosmology with total matter density $\Omega_{\rm m}=0.272$, dark energy density $\Omega_\Lambda=0.728$, amplitude of the matter power spectrum $\sigma_8=0.81$, baryon density $\Omega_{\rm b}=0.045$, Hubble constant $H_0=70.4 \, \rm km\,s^{-1}\,Mpc^{-1}$, and $n_{\rm s}=0.967$ compatible with the Wilkinson Microwave Anisotropy Probe 7 cosmology~\citep{komatsuetal11}.
The values of this set of cosmological parameters are compatible with those of the recent \emph{Planck} results within a 10 per cent relative variation~\citep{plancketal13cosmo}.
The size of the box is $L_{\rm box}=100 \, h^{-1}\rm\,Mpc$ with $1024^3$ DM particles, which results in a DM mass resolution of $M_{\rm DM, res}=8\times 10^7 \, \rm M_\odot$.
The initial conditions have been produced with the {\sc mpgrafic} software~\citep{prunetetal08}.
The simulation was run down to $z=1.2$ and used 4 million CPU hours.

The \hagn simulation is run with the adaptive mesh refinement code {\sc ramses}~\citep{teyssier02}.
The evolution of the gas is followed using a second-order unsplit Godunov scheme for the Euler equations. 
The HLLC Riemann solver~\citep{toroetal94} with MinMod total variation diminishing scheme is used to reconstruct the interpolated variables from their cell-centred values. 
Collisionless particles (DM and star particles) are evolved using a particle-mesh solver with a cloud-In-cell interpolation.
The initial mesh is refined up to $\Delta x=1\, \rm kpc$ (seven levels of refinement).
This is done according to a quasi-Lagrangian criterion: if the number of DM particles in a cell is more than 8, or if the total baryonic mass in a cell is eight times the initial DM mass resolution, a new refinement level is triggered.
In order to keep the minimum cell size approximately constant in physical units, we allow a new maximum level of refinement every time the expansion scale factor doubles (i.e. at $a_{\rm exp}=0.1$, $0.2$, $0.4$ and $0.8$).

\subsubsection{Gas cooling and heating}

Gas is allowed to cool by H and He cooling with a contribution from metals using a~\cite{sutherland&dopita93} model down to $10^4\, \rm K$. 
Heating from a uniform UV background takes place after redshift $z_{\rm reion} = 10$ following~\cite{haardt&madau96}. 
Metallicity is modelled as a passive variable for the gas and its amount is modified by the injection of gas ejecta during supernova (SN) explosions and stellar winds.
We also account for the release of various chemical elements synthesized in stars and released by stellar winds and SNe: O, Fe, C, N, Mg and Si. 
However, they do not contribute separately to the cooling curve (the ratio between each element is taken to be solar for simplicity) but can be used to probe the distribution of the various metal elements.
The gas follows an equation of state for an ideal monoatomic gas with an adiabatic index of $\gamma=5/3$.

\subsubsection{Star formation and stellar feedback}

The star formation process is modelled with a Schmidt law:
$\dot \rho_*= \epsilon_* {\rho / t_{\rm ff}}\, ,$ where $\dot \rho_*$ is the star formation rate (SFR) density, $\epsilon_*=0.02$~\citep{kennicutt98, krumholz&tan07} the constant star formation efficiency and $t_{\rm ff}$ the local free-fall time of the gas.
Star formation is allowed in regions which exceed a gas hydrogen number density threshold of $n_0=0.1\, \rm H\, cm^{-3}$ following a Poissonian random process~\citep{rasera&teyssier06, dubois&teyssier08winds} with a stellar mass resolution of $M_*=\rho_0 \Delta x^3\simeq 2\times 10^6 \, \rm M_\odot$.
The gas pressure is artificially enhanced above $\rho > \rho_0$ assuming a polytropic equation of state $T=T_0(\rho/\rho_0)^{\kappa-1}$ with polytropic index $\kappa=4/3$ to avoid excessive gas fragmentation and mimic the effect of stellar heating on the mean temperature of the interstellar medium~\citep{springel&hernquist03}.
Feedback from stars is explicitly taken into account assuming a \citet{salpeter55} initial mass function (IMF) with a low-mass (high-mass) cut-off of $0.1\, \rm M_{\odot}$ ($100 \, \rm M_{\odot}$), as described in detail in Kimm et al. (in preparation). 
Specifically, the mechanical energy from Type II SNe and stellar winds is taken from {\sc starburst99}~\citep{leithereretal99, leithereretal10}, and the frequency of Type Ia SN explosions is computed following~\cite{greggio&renzini83}. 

\subsubsection{Feedback from black holes}

The same `canonical'  active galactic nuclei (AGN) feedback modelling employed in~\cite{duboisetal12agnmodel} is used here.
Black holes (BHs) are created where the gas mass density is larger than $\rho > \rho_0$ with an initial seed mass of $10^5\, \rm M_\odot$.
In order to avoid the formation of multiple BHs in the same galaxy, BHs are not allowed to form at distances less than 50 kpc from each other.
The accretion rate on to BHs follows the Bondi--Hoyle--Lyttleton rate
$\dot M_{\rm BH}=4\pi \alpha G^2 M_{\rm BH}^2 \bar \rho / (\bar c_{\rm s}^2+\bar u^2) ^{3/2},$
where $M_{\rm BH}$ is the BH mass, $\bar \rho$ is the average gas density, $\bar c_{\rm s}$ is the average sound speed, $\bar u$ is the average gas velocity relative to the BH velocity, and $\alpha$ is a dimensionless boost factor with $\alpha=(\rho/\rho_0)^2$ when $\rho>\rho_0$ and $\alpha=1$ otherwise~\citep{booth&schaye09} in order to account for our inability to capture the colder and higher density regions of the interstellar medium.
The effective accretion rate on to BHs is capped at the Eddington accretion rate:
$\dot M_{\rm Edd}=4\pi G M_{\rm BH}m_{\rm p} / (\epsilon_{\rm r} \sigma_{\rm T} c),$
where $\sigma_{\rm T}$ is the Thompson cross-section, $c$ is the speed of light, $m_{\rm p}$ is the proton mass and $\epsilon_{\rm r}$ is the radiative efficiency, assumed to be equal to $\epsilon_{\rm r}=0.1$ for the \cite{shakura&sunyaev73} accretion on to a Schwarzschild BH.

The AGN feedback is a combination of two different modes, the so-called \emph{radio} mode operating when $\chi=\dot M_{\rm BH}/\dot M_{\rm Edd}< 0.01$ and the \emph{quasar} mode active otherwise.
The quasar mode consists of an isotropic injection of thermal energy into the gas within a sphere of radius $\Delta x$, and at an energy deposition rate: $\dot E_{\rm AGN}=\epsilon_{\rm f} \epsilon_{\rm r} \dot M_{\rm BH}c^2$. In this equation, $\epsilon_{\rm f}=0.15$ is a free parameter chosen to reproduce the scaling relations between BH mass and galaxy properties (mass, velocity dispersion) and BH density in our local Universe (see \citealp{duboisetal12agnmodel}).
At low accretion rates, the radio mode deposits AGN feedback energy into a bipolar outflow with a jet velocity of $10^4\,\rm km\, s^{-1}$.
The outflow is modelled as a cylinder with a cross-sectional radius $\Delta x$ and height $2 \, \Delta x$ following~\cite{ommaetal04} (more details  are given in~\citealp{duboisetal10}).
The efficiency of the radio mode is larger than the quasar mode with $\epsilon_{\rm f}=1$.

A projected map of half the simulation volume and a smaller subregion is shown in Fig.~\ref{fig:h-agn}. 
Gas density, gas temperature and gas metallicity are depicted.
One can discern the large-scale pattern of the cosmic web, with filaments and walls surrounding voids and connecting haloes. Massive haloes are filled with hot gas, and feedback from SNe and AGN pours warm and metal-rich gas in the diffuse inter-galactic medium.
As demonstrated in \cite{Dubois2013}, the modelling of AGN feedback is critical to create early-type galaxies and provide the sought morphological diversity (see Fig.~\ref{fig:h-agn_gal} for a snippet of the galaxy sample of the simulation) in hydrodynamical cosmological simulations~\citep[see e.g.][for semi-analytical models]{crotonetal06}.

\subsection{Mock observations of galaxies}
\label{section:postprocess}

We describe how we produce various observables that can be compared qualitatively with data from modern observational surveys. In this paper we focus on observables which are known to correlate with the Hubble type of galaxies, namely mass, $V/\sigma$, colour, morphological parameters like Gini and $M_{20}$, and age.

\subsubsection{Identifying  and segmenting galaxies}

Galaxies are identified with the AdaptaHOP finder (\citealp{aubertetal04}, updated to its recent version by~\citealp{tweedetal09} for building merger trees) which directly operates on the distribution of star particles. A total of 20 neighbours are used to compute the local density of each particle, a local threshold of $\rho_{\rm t}=178$ times the average total matter density is applied to select relevant densities, and the force softening (minimum size below which substructures are considered irrelevant) is $\sim2$ kpc. Only galactic structures identified with more than 50 particles are considered. It allows for a clear separation of galaxies (defined as sets of star particles segmented by AdaptaHOP), including those in the process of merging. Catalogues of around $\sim 150 \, 000$ galaxies are produced for each redshift analysed in this paper from $z=3$ to $1.2$.

\subsubsection{Synthetic colours}

We compute the absolute AB magnitudes and rest-frame colours of galaxies using single stellar population models from~\cite{bruzual&charlot03} assuming a Salpeter IMF. Each star particle contributes to a flux per frequency that depends on its mass, age, and metallicity. The sum of the contribution from all stars is passed through the \emph{u}, \emph{g}, \emph{r} and \emph{i} filters from the SDSS. Fluxes are expressed as rest-frame quantities (i.e. that do not take into account the red-shifting of spectra). We also neglect the contribution to the reddening of spectra from internal (interstellar medium) or external (intergalactic medium) dust extinction.
Once the flux in each waveband is obtained for a star particle, we build two-dimensional projected maps from single galaxies (satellites are excised with the galaxy finder), and we can sum up the total contribution of their stars to the total luminosity.
A small sample of galaxies representative of the morphological variety in the simulation is shown in Fig.~\ref{fig:h-agn_gal}.

\subsubsection{Projected stellar kinematics}

\begin{figure}
   \includegraphics[width=0.495\columnwidth]{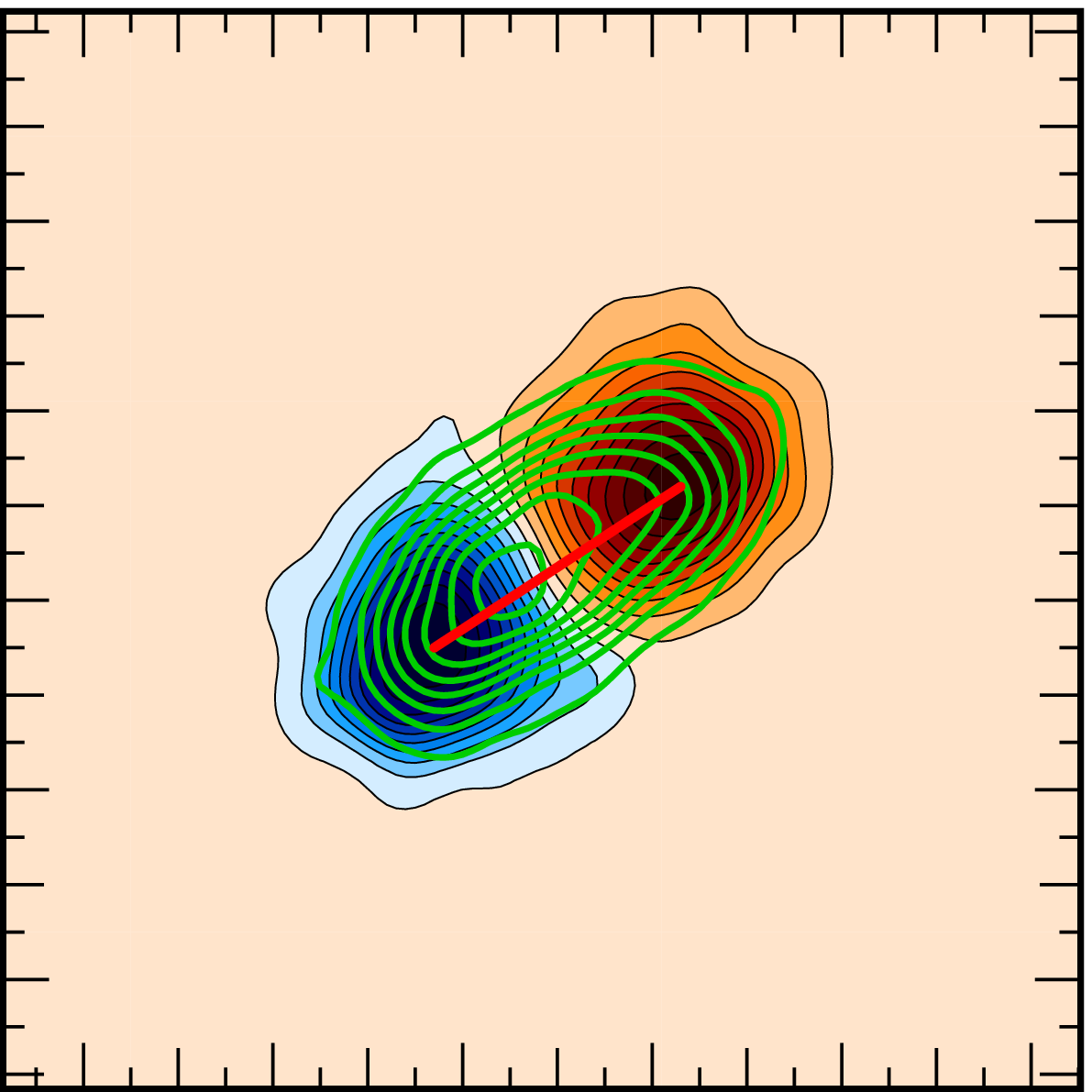}   
    \includegraphics[width=0.495\columnwidth]{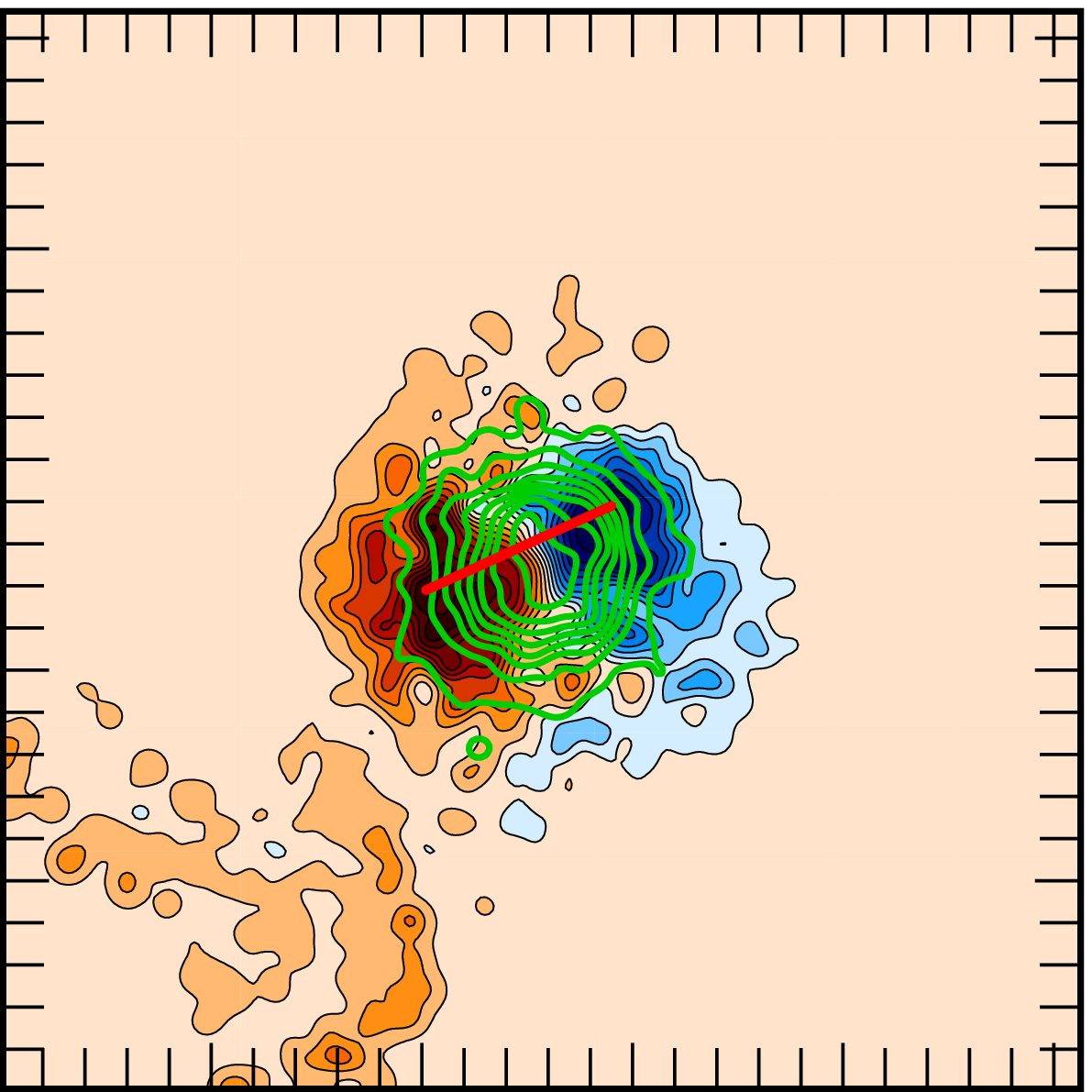}   
        \caption{Example of integral field spectroscopy for the velocity of a fast rotator $V/\sigma>1$ (left-hand panel) and a slow rotator $V/\sigma<1$ (right-hand panel). The thick red line corresponds to the position of the slit, which is placed along the kinetic major axis. The green iso-contours correspond to the velocity dispersion map. The size of the images is 50 kpc and the mean velocity (respectively dispersion) amplitude is 150 (respectively 75) $\rm km\, s^{-1}$ (left-hand panel) and 50 (respectively 100)  $\rm km\, s^{-1}$ (right-hand panel).}
\label{fig:IFU}
\end{figure}

For each galaxy, we build a field of view centred on the galaxy, which is made of 256$\times$256 pixels over 100 kpc size (corresponding to a pixel size of $0.4$ kpc or $0.05$ arcsec at $z=1.83$).
We compute the luminosity-weighted velocity along the line of sight (arbitrary defined as the \emph{x}-axis of the simulation):
\begin{equation}
\bar v_{\rm pixel}= { \Sigma_{\rm i} v_{\rm los, i} I_{\rm i, filter} \over \Sigma_{\rm i} I_{\rm i, filter} }\, ,
\end{equation}
where $v_{\rm los, i}$ is the velocity along the line of sight of the \emph{i}-th star in the pixel considered and $I_{\rm i, filter}$ is the intensity in the corresponding filter bandwidth (\emph{u, g, r, i}) of the \emph{i}-th star in the pixel considered. 
Then, the velocity dispersion along the line of sight is:
\begin{equation}
\bar \sigma_{\rm pixel}^2= { \Sigma_{\rm i} v_{\rm los, i}^2 I_{\rm i, filter} \over \Sigma_{\rm i} I_{\rm i, filter} } - \bar v_{\rm pixel}^2 \, .
\end{equation}
The velocity maps are then smoothed with a Gaussian kernel of 15 pixels.
The position of the fastest (respectively slowest) pixel, which defines $V$ for that galaxy, is then identified automatically and a $0.75$ arcsec ``slit'' is put across so as to interpolate through the kinematic major axis of the galaxy. 
The smoothed velocity dispersion map is also interpolated along the same axis and the maximum of that curve defines $\sigma$ (see Fig.~\ref{fig:IFU} for example of a slow and a fast rotator).
$V/\sigma$ is then straightforwardly the corresponding ratio.

\subsubsection{Specific star formation rate}

The calculation of the SFR is done on stars as identified by the galaxy finder that belong to a given galaxy.
To compute the SFR, we compute the amount of stars formed over the last 100 Myr.
The choice of $100 \, \rm Myr$ corresponds to a minimum measurable SFR of $M_*/100 \,\rm Myr=0.02 \, \rm M_\odot\, yr^{-1}$.
The specific star formation rate (sSFR) is then calculated by diving the SFR by the galaxy stellar mass, $M_{\rm s}$. 

\subsubsection{Gini and $M_{20}$}

The morphology of each galaxy is often measured by two non-parametric parameters: the Gini~\citep[$G$,][]{abrahametal03} and $M_{20}$ \citep{lotzetal04}. The Gini parameter is a non-parametric measure of the inequality of fluxes in pixels, ranging from zero (for a perfectly uniform image) to unity (for an image with all the flux in one pixel, for instance). $M_{20}$ is the second order momentum of light of the 20 per cent brightest pixels of a galaxy. It traces the spatial distribution of any bright nuclei, bars, spiral arms, and off-centre star clusters. As shown in \citet{lotzetal04}, galaxies with $M_{20} > -1.1$ mainly are extended objects with double or multiple nuclei, whereas low values of $M_{20}$ ( $< -1.6$) are relatively smooth with a bright nucleus. Both parameters are known to correlate well with the concentration parameter \citep{abrahametal94} for regular shapes, but they are better suited for disturbed morphologies because of their non-parametric nature. These 
two parameters have been used to characterize observations in the local universe and at high redshift \citep[e.g.][]{lotzetal04, abrahametal07, wangetal12, leeetal13} and they are well suited to analyse large samples of galaxies of mixed morphologies. In the local universe, galaxies with a high Gini value and a low $M_{20}$ value are mainly ellipticals, whereas late-type galaxies and irregular have lower $G$  and larger $M_{20}$ values. Mergers tend to have large $G$ and large $M_{20}$ values.

Images in the $i$ band are obtained from a segmentation of 3D objects with the galaxy finder. The images are rebinned to 64$\times$64 pixels for a 100 kpc size image in order to avoid star particles appearing as individual pixels. Then, as in \citet{leeetal13}, we measure the Petrosian radius with an elliptical aperture which is obtained as in {\sc SExtractor}~\citep{bertinetal96} from the second-order moment of light. The Petrosian semi-major axis $a_{\rm p}$ is such that the ratio of the surface brightness at $a_{\rm p}$ over the mean surface brightness within $a_{\rm p}$ is decreasing at $a_{\rm p}$ and becomes smaller than 0.2. In practice, we fit a spline to the surface brightness ratio profile and find the zero of the function $\mu(a_{\rm p})/\mu(<a_{\rm p})-0.2$.
Galaxies with $a_{\rm p}$ smaller than 2 pixels are filtered out: they are almost always associated with low-mass galaxies ($M_{\rm s}<10^{9.2}\, \rm M_\odot$) with few star particles, and the $G$ and $M_{20}$ parameters are very uncertain for these objects. We also filter out galaxies less massive than $M_{\rm s}<10^{9.5} \, \rm M_\odot$ for which we suffer the most from resolution effects.
A description of the bivariate distributions of $G$, $M_{20}$ and stellar mass is given in Appendix~\ref{sec:morpho}.

\subsubsection{Ages}

The mean ages of galaxies are obtained through the summation of the mass-weighted age of star particles belonging to the galaxy.

\subsubsection{Spin of galaxies}

To compute the spin of galaxies, we compute the total angular momentum of their stars with respect to the particle of maximum density (centre of the galaxy) from the smoothed stellar density constructed with the AdaptaHOP algorithm.

We have also tested the effect of grid-locking on the Cartesian axes of the box (a common issue of Cartesian-based Poisson solvers for which a numerical anisotropy in the force calculation arises;~\citealp[see e.g.][]{hockney&eastwood81}) in Appendix~\ref{sec:spinlock} for galaxies and filaments.

\begin{figure}
  \hskip 0.3cm \includegraphics[width=0.95\columnwidth]{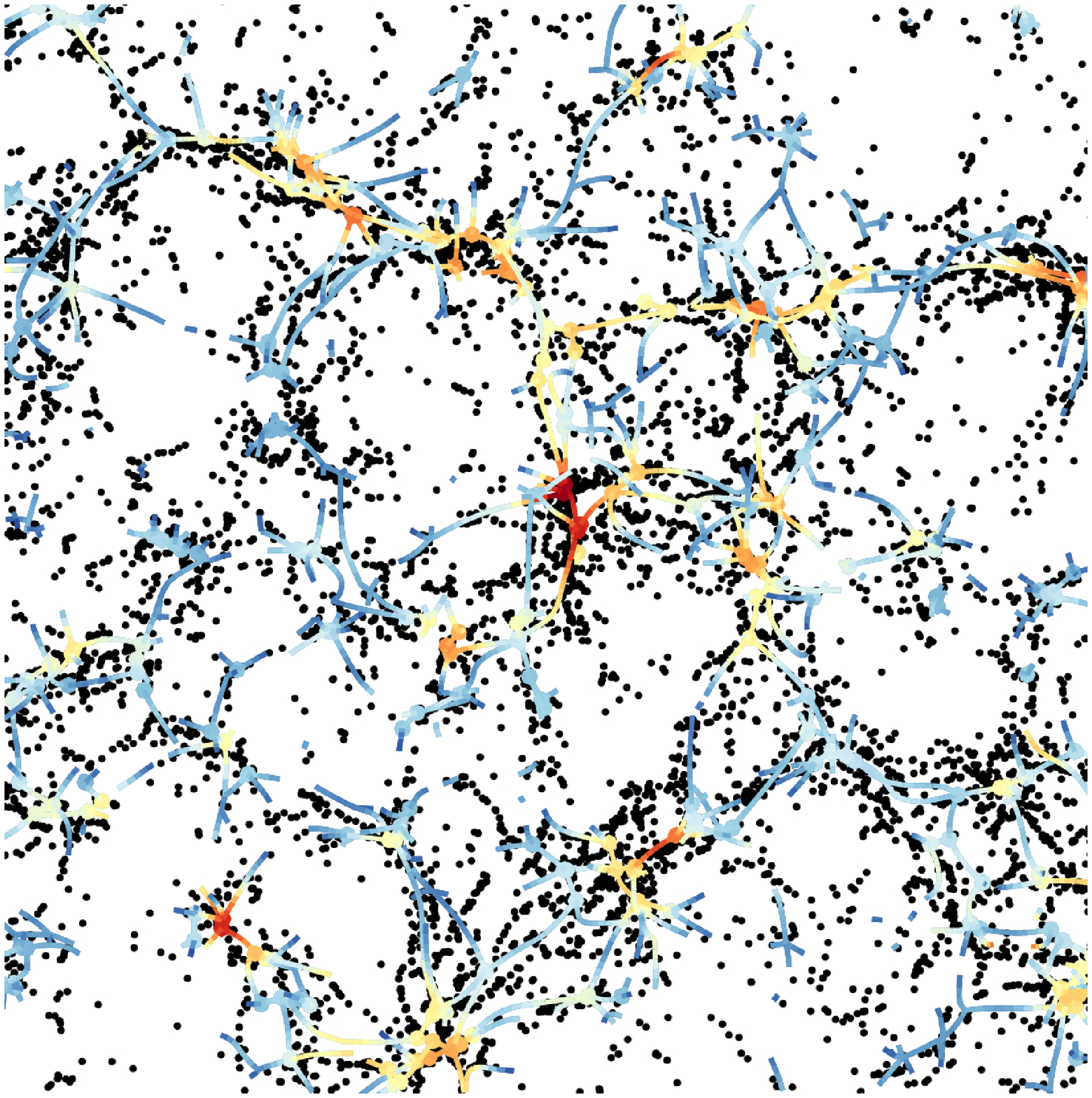}
  \hskip 0.45cm \includegraphics[width=0.975\columnwidth]{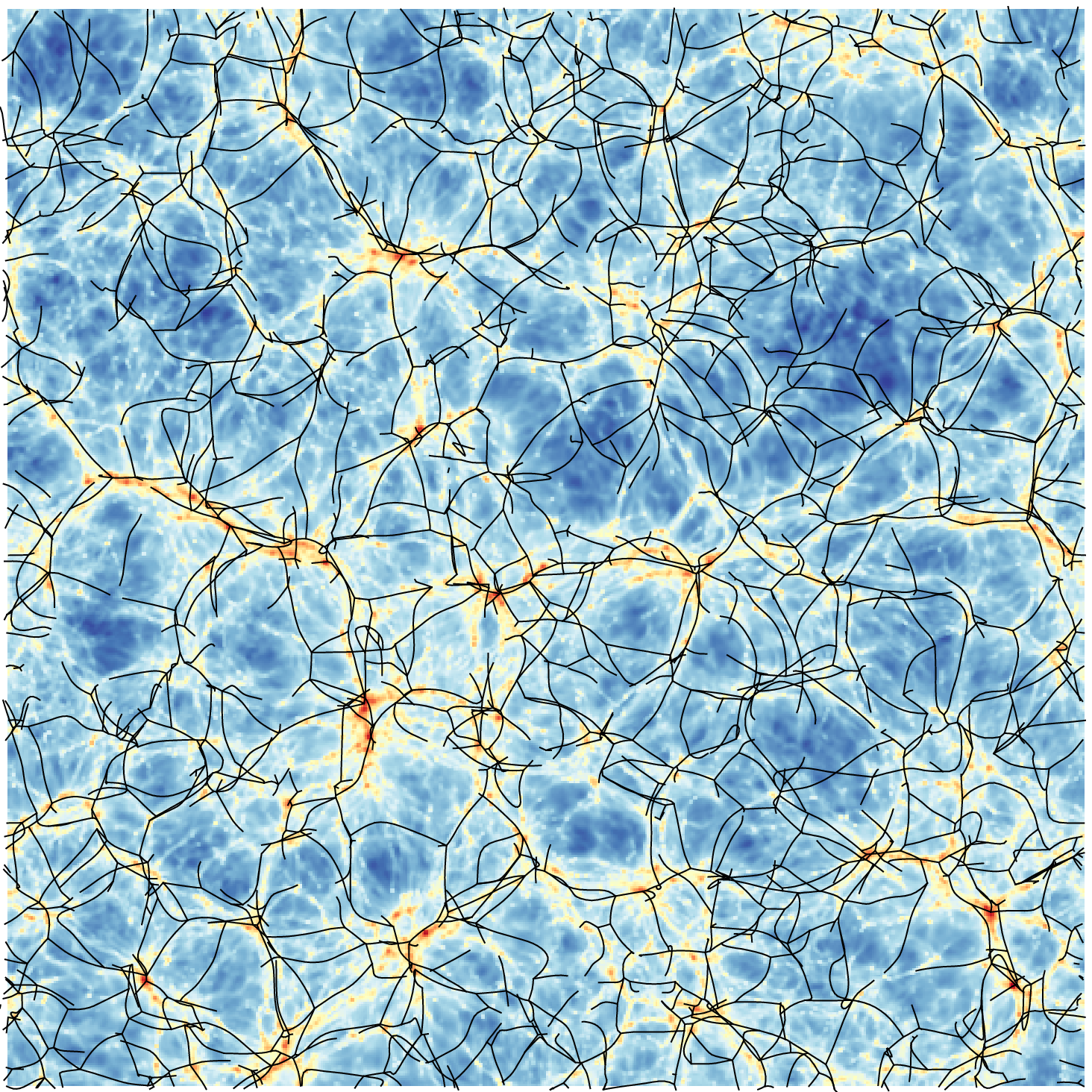}
        \caption{Top: projection along the $z$-axis of the \hagn gas skeleton (colour coded by logarithmic density as red--yellow--blue--white from high density to low density) at redshift $z=1.83$ of a slice of $25\, h^{-1}\, \rm Mpc$ on the side and $10\, h^{-1}\, \rm Mpc$ thickness. Galaxies are superimposed as black dots. The clustering of the galaxies follows the skeleton quite closely. Bottom: larger view of the skeleton on top of the projected gas density. This paper quantifies orientation of the galaxies relative to the local anisotropy set by the skeleton. 
        }
\label{fig:skeleton}
\end{figure}

\subsection{Tracing large-scale structures via the skeleton}
\label{section:skeleton}

In order to quantify the orientation of galaxies relative to the cosmic web, we use a geometric three-dimensional ridge extractor well suited to identify filaments, called the `skeleton'.
A gas density cube of $512^3$ pixels is drawn from the simulation and Gaussian-smoothed with a length of $3 \, h^{-1}\, \rm Mpc$ comoving chosen so as to trace large-scale filamentary features.
Two implementations of the skeleton, based on `watershed'~\citep{sousbie09} and `persistence'~\citep{sousbie10}, were implemented, without significant difference for the purpose of this investigation.
The first method identifies ridges as the boundaries of walls which are themselves the boundaries of voids.
The second one identifies ridges as the `special' lines connecting topologically robust (filament-like) saddle points to peaks. 
 
Fig.~\ref{fig:skeleton} shows a slice of $25 \, h^{-1}\, \rm Mpc$ of the skeleton colour coded by logarithmic density, along with galaxies contained within that slice. 
The clustering of the galaxies follows quite closely the skeleton of the gas, i.e. the cosmic filaments.
Note that, on large scales, the skeleton built from the gas is equivalent to that built from the DM as the gas and DM trace each other closely.
The rest of the paper is devoted to studying  the orientation of the spin of these galaxies relative to the direction of the nearest skeleton segment.
In practice, an octree is built from the position of the mid-segment of the skeleton to speed up the association of the galaxy position to its nearest skeleton segment.
It was checked that our results were not sensitive to  how many such segments were considered to define the local direction of the skeleton.
The orientation of the segment of the skeleton is used to define the relative angle between the filament and the spin of the galaxy.
The segments are also tagged with their curvilinear distance to the closest node (where different filaments merge), which allows us to study the evolution of this (mis)alignment \emph{along} the cosmic web.
Appendix~\ref{sec:spinlock} investigates the effect of grid-locking of the skeleton's segments in the \hagn simulation.
Large-scale filaments, defined from the skeleton, do not show any alignment with the grid. 

\begin{figure*}
   \includegraphics[width=5.5cm]{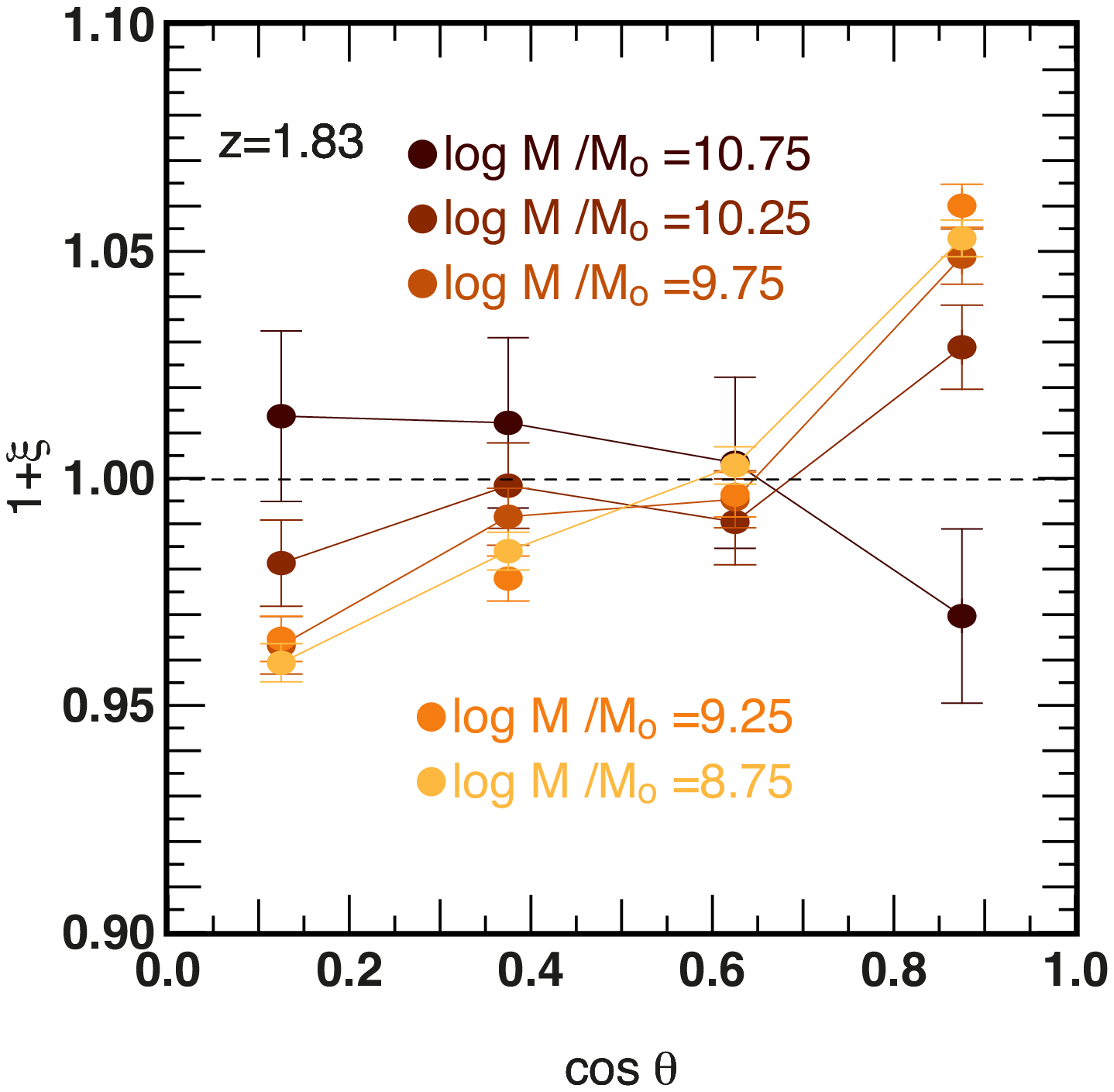}   
   \includegraphics[width=5.5cm]{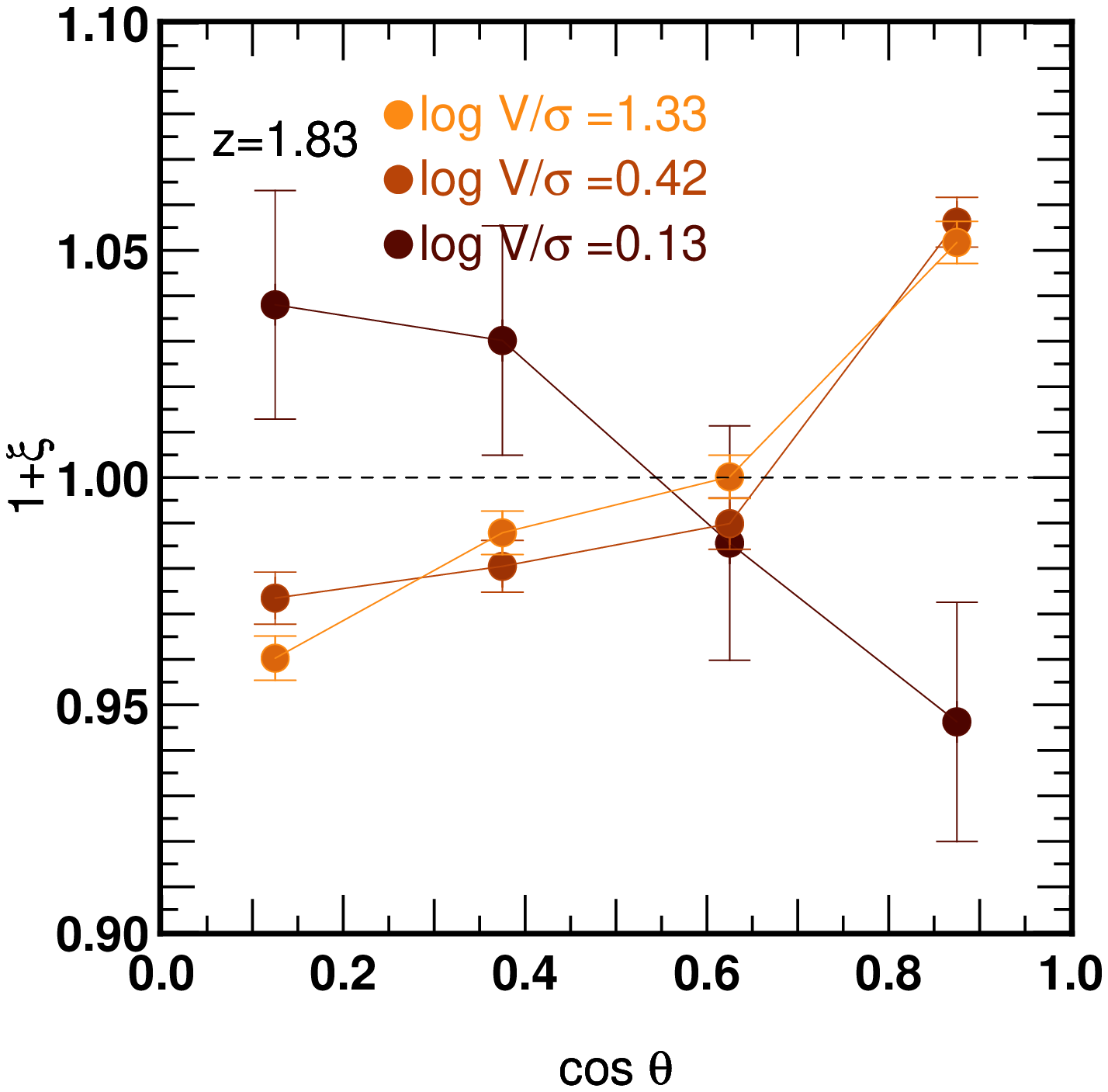}   
  \includegraphics[width=5.5cm]{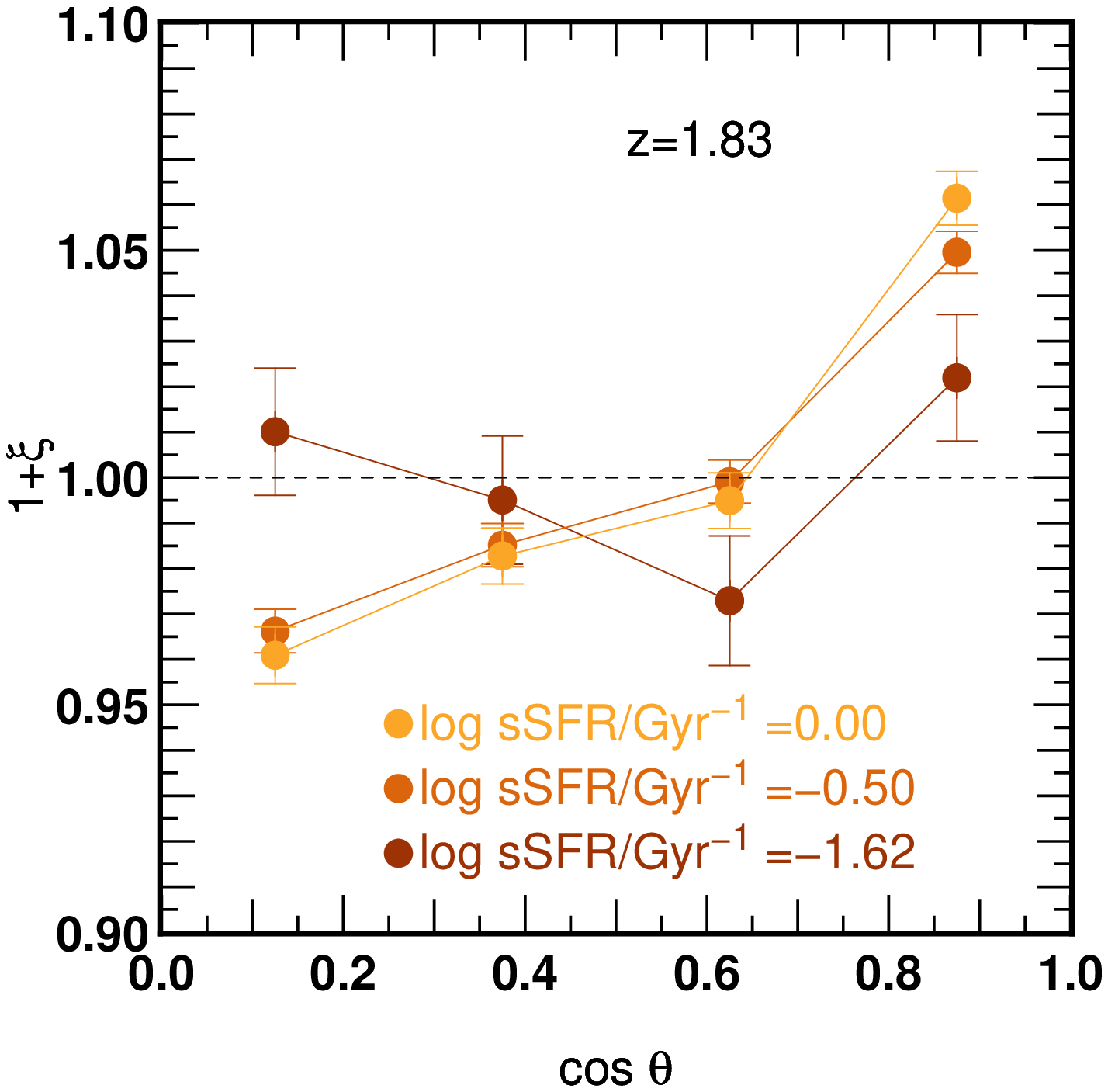}
   \includegraphics[width=5.5cm]{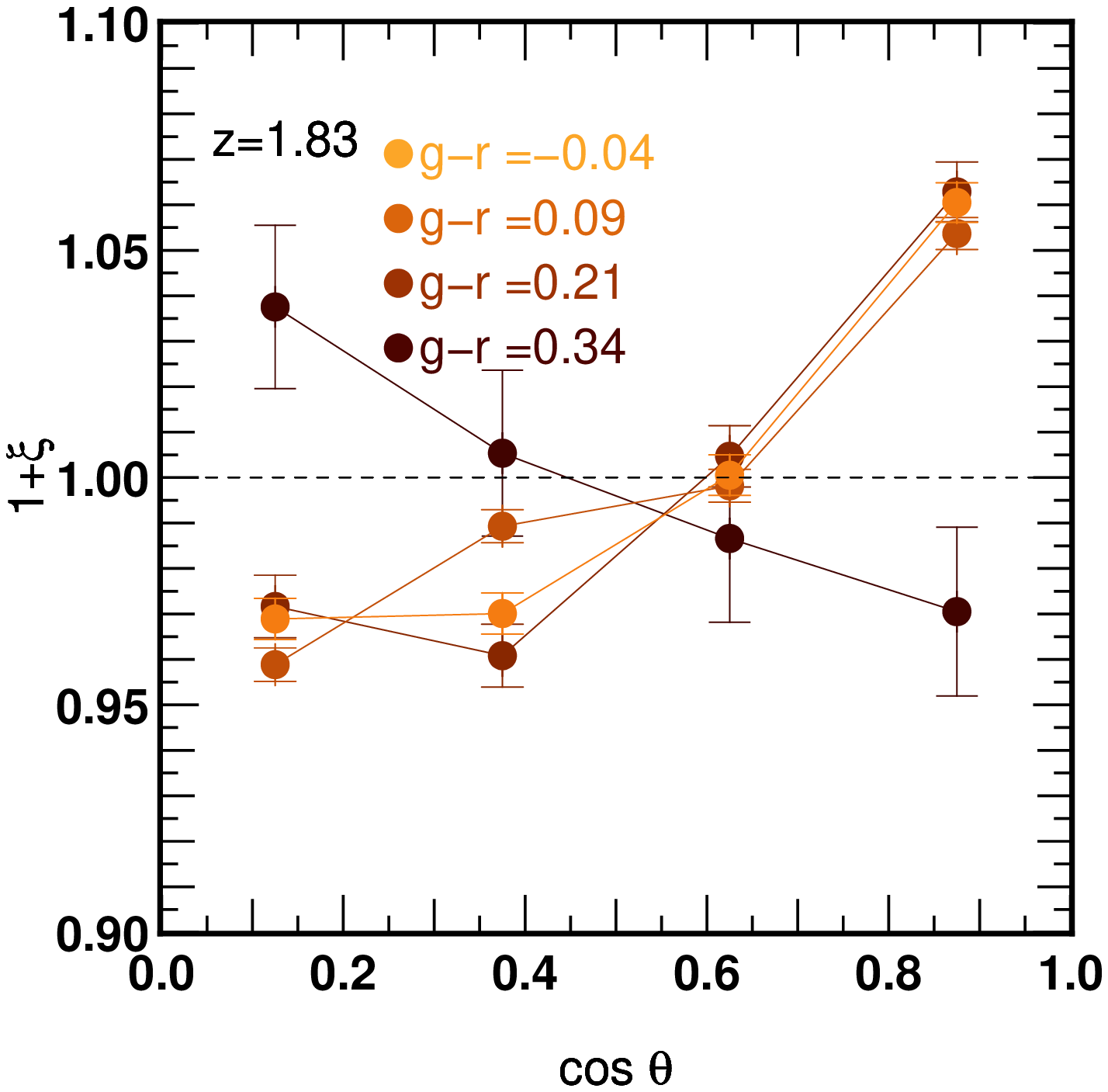}
   \includegraphics[width=5.5cm]{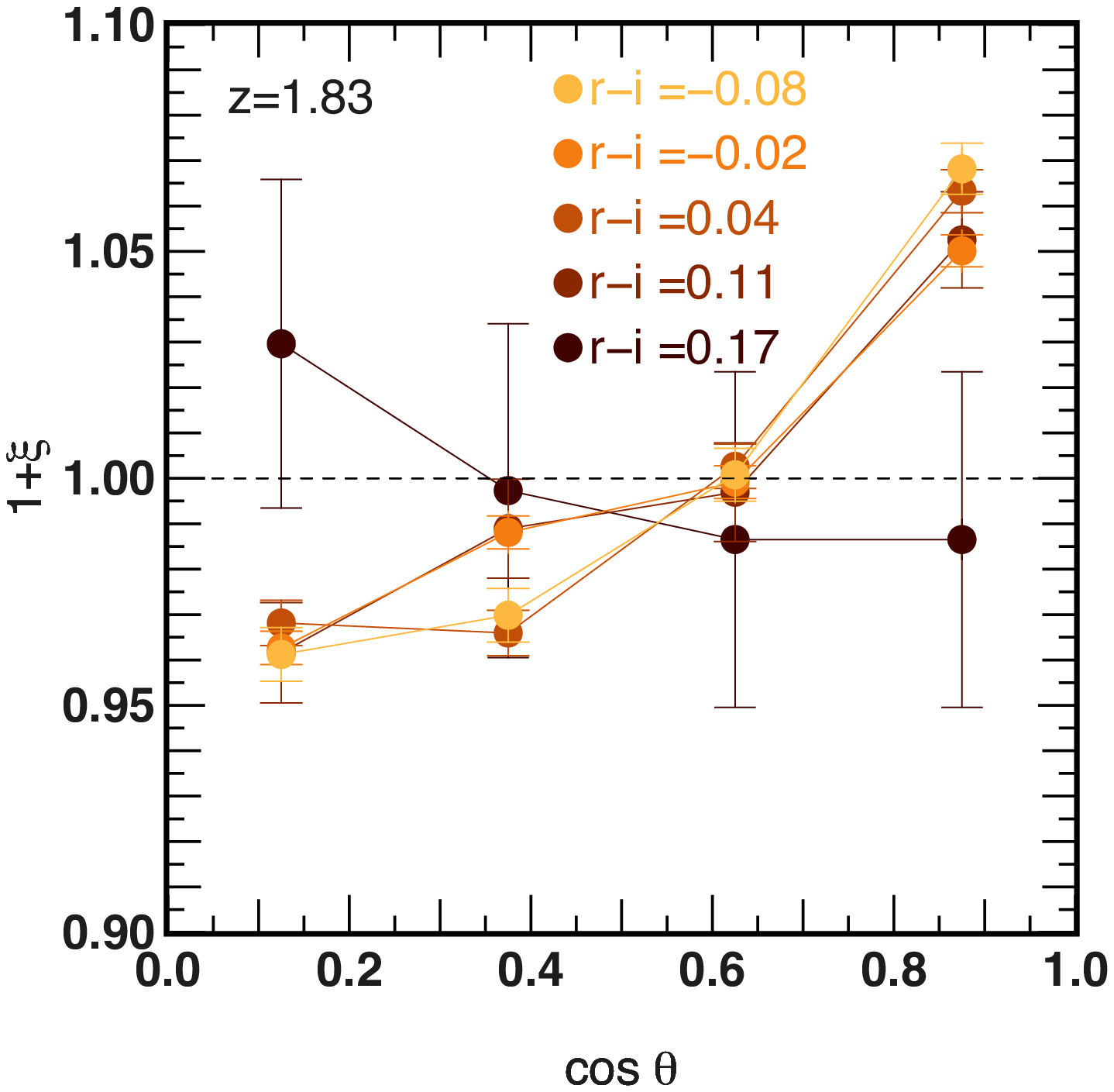}   
 \includegraphics[width=5.5cm]{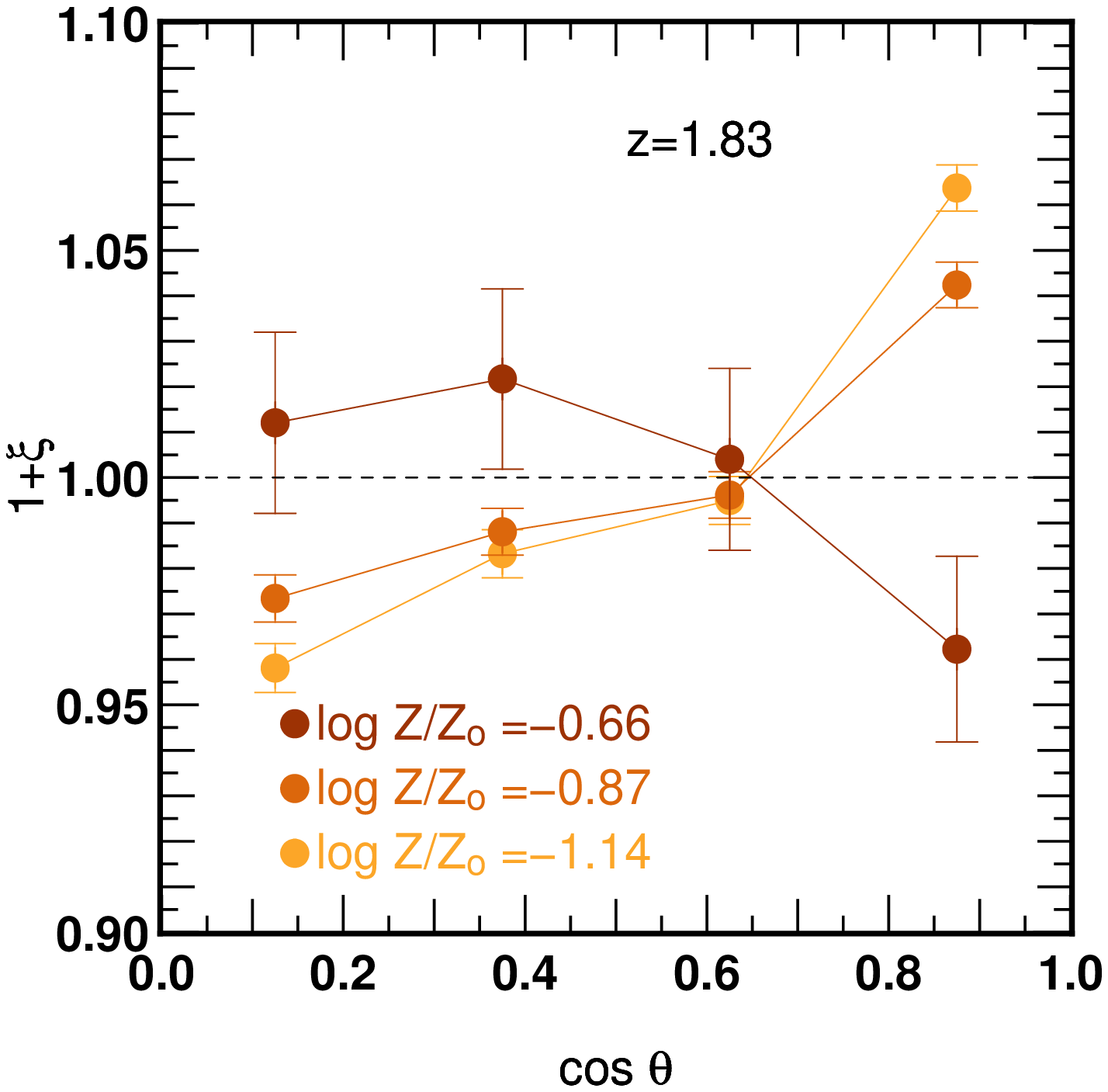}
 \includegraphics[width=5.5cm]{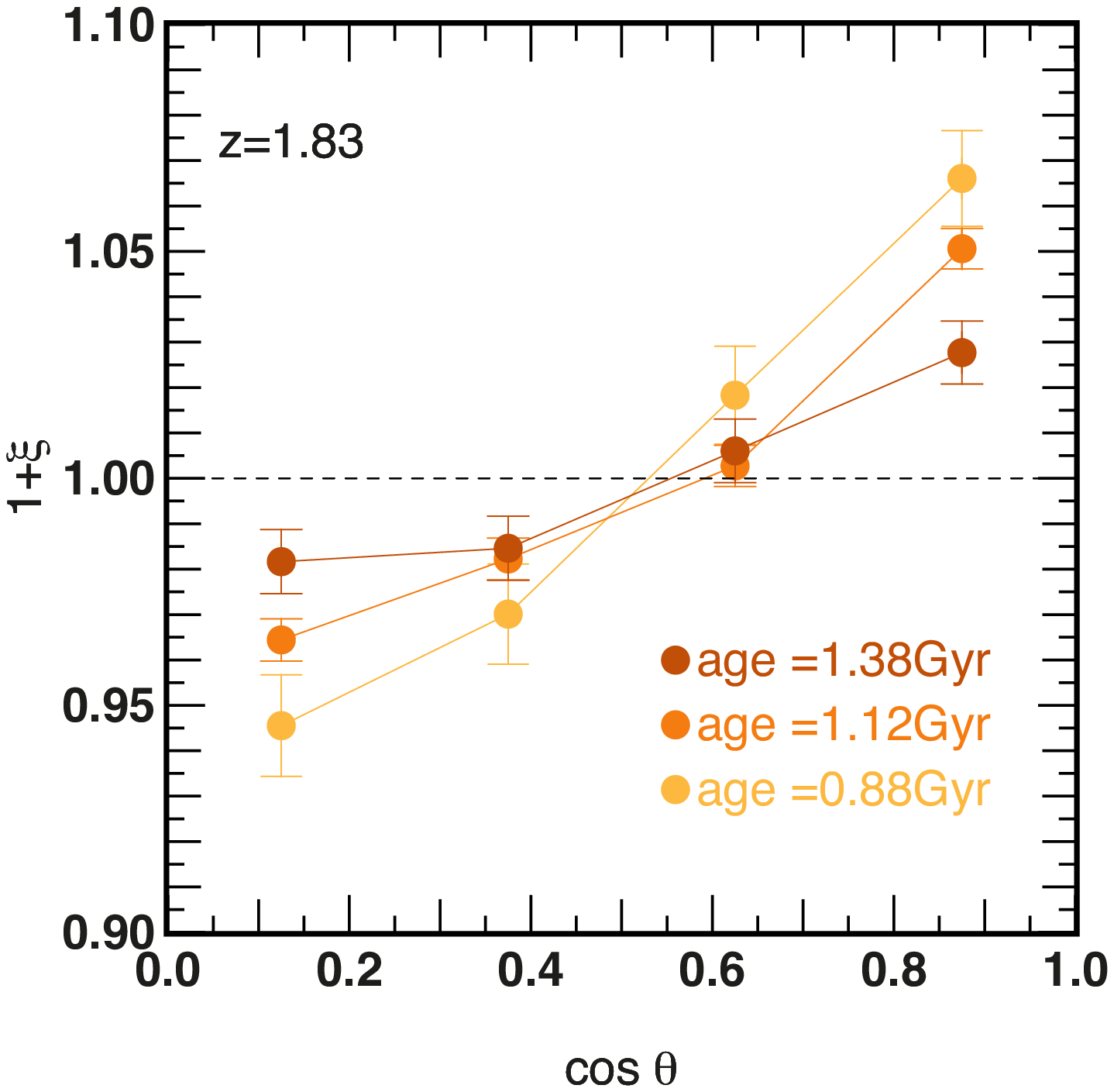}
 \includegraphics[width=5.5cm]{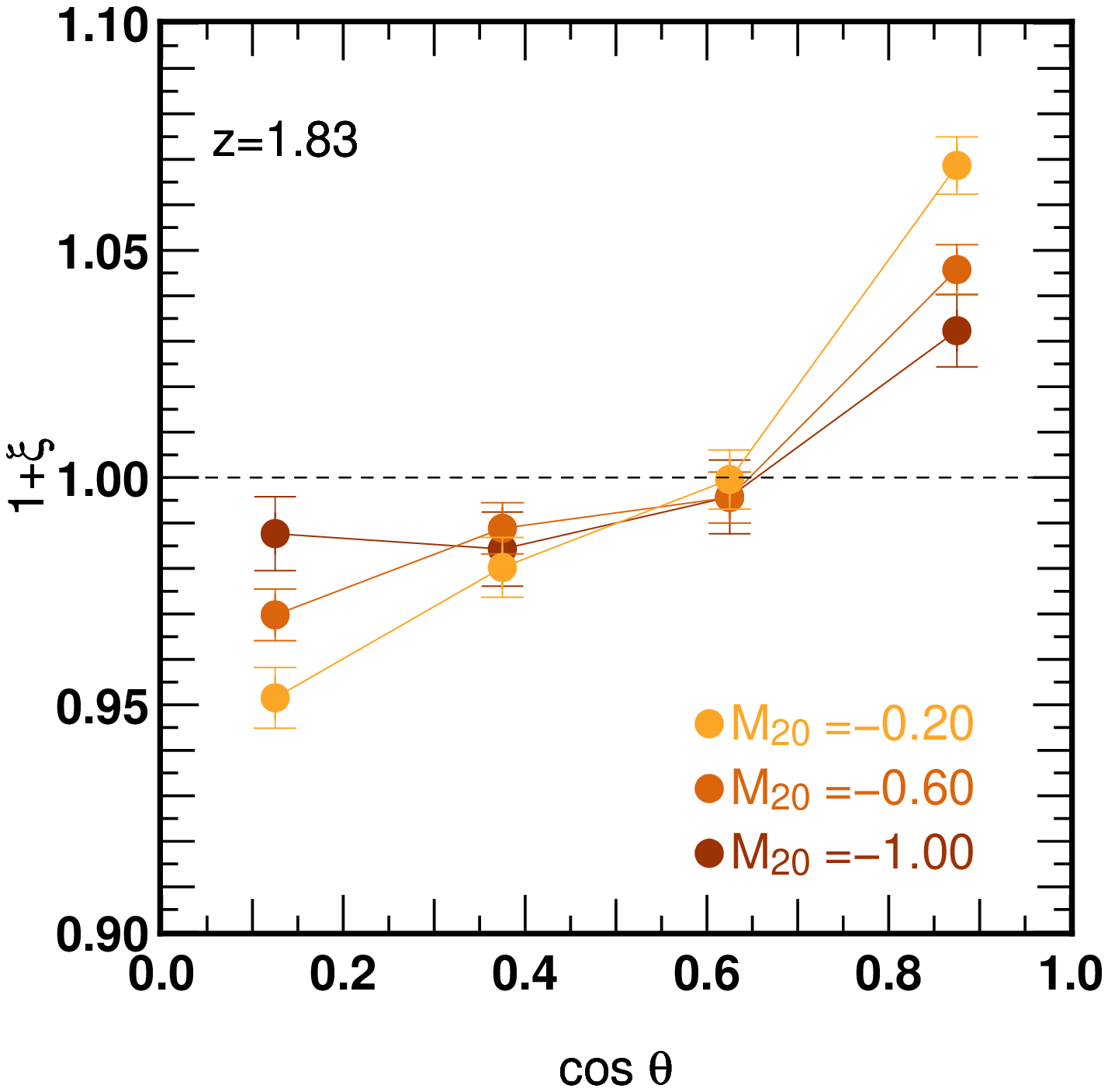}
 \includegraphics[width=5.5cm]{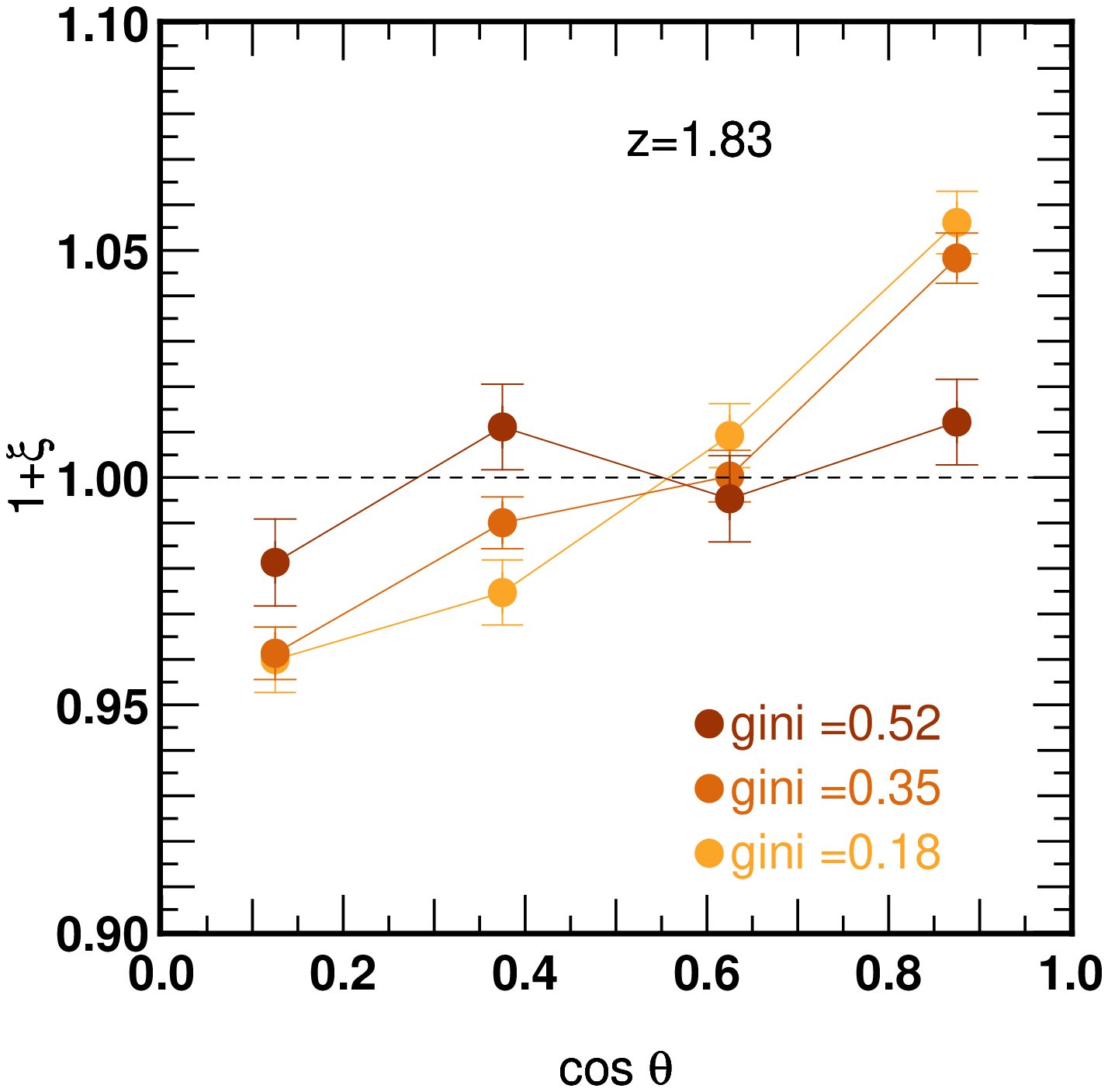}
        \caption{Excess probability, $\xi$, of the alignment between the spin of galaxies and their closest filament is shown as a function of galaxy properties at $z=1.83$: $M_{\rm s}$ (top row, left column), $V/\sigma$ (top row, middle column), sSFR (top row, right column), $g-r$ (middle row, left column), $r-i$ (middle row, middle column), metallicity $Z$ (middle row, right column), age (bottom row, left column), $M_{\rm 20}$ (bottom row, middle column) and Gini (bottom row, right column). 
Half-sigma error bars are shown for readability. Dashed line is uniform PDF (excess probability $\xi=0$).
Massive, dispersion-dominated, passive, red, smooth and old galaxies tend to have a spin perpendicular or randomly oriented with the direction of their filament. Low-mass, centrifugally supported, star-forming, blue, irregular and young galaxies tend to align with the direction of their closest filament.}
\label{fig:fullxi}
\end{figure*}

\section{Spin swing evolution}
\label{sec:result}

This paper focuses on the orientation of the spin of galaxies relative to the filaments in which they are embedded and the cosmic evolution.
Specifically, we aim to see if its evolution can be traced via physical and morphological tracers.
We investigate in Section~\ref{sec:orient} how this orientation varies with different tracers of the Hubble type of galaxies, namely stellar mass, $V/\sigma$, sSFR, colour, metallicity, age,  $M_{20}$ and Gini on our sample of 150 000 galaxies. 
Afterwards, in Section~\ref{sec:along}, we quantify how this alignment varies as a function of distance to the filaments and along the filaments to the nodes of the cosmic web.
We study the cosmic evolution of the alignment of the spin of galaxies and filaments in Section~\ref{sec:evolution}.

\subsection{Alignment of galaxies and filaments}
\label{sec:orient}

We measure the statistical signature of the (mis)alignment of galaxies with their closest filament segment.
The alignment is defined as the angle $\theta$ between the spin of the stellar component and the direction of the filamentary segment.
Recall that the spin of the galaxy is obtained by removing merging substructures using the galaxy finder, and computing the net angular momentum of its stars with respect to its centre (defined as the point of highest stellar density).
Note that the filament segments are assumed to have no polarity. 
Hence we impose the angle $\theta$ has a $\pi/2$ symmetry, and is expressed in terms of $\cos \theta=[0,1]$.

Fig.~\ref{fig:fullxi} shows the resulting probability density function (PDF), $1+\xi$, at $z=1.83$, where $\xi$ is the excess probability of $\cos \theta$ in bins of  various quantities: mass, kinematics, sSFR, colour, metallicity, age, $M_{\rm 20}$ and Gini. 
A uniform PDF (i.e. random orientations of galaxies relative to their filament) is represented as a dashed line for comparison.
Galaxies with mass below $M_{\rm s}< 10^9\, \rm M_\odot$ are removed from the calculation, except for investigation of alignment as a function of mass. 

\begin{figure}
\center\includegraphics[width=0.85\columnwidth]{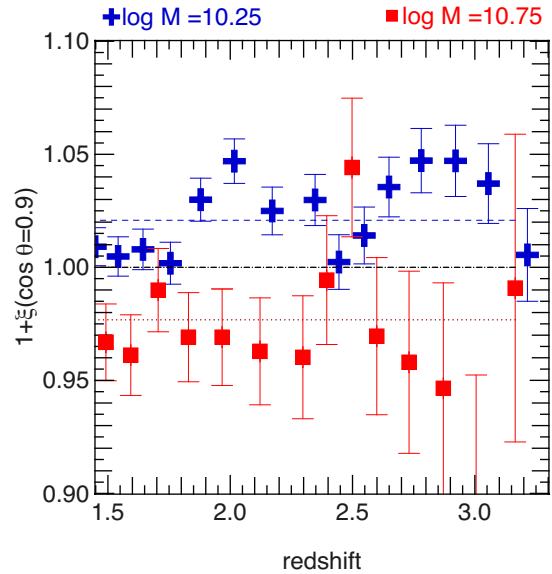}
\caption{Average values of $1+\xi(\cos \theta =0.9)$ as a function of redshift for two different bins of stellar mass $M_{\rm s}=10^{10.25}$ (pluses) and $M_{\rm s}=10^{10.75}$ (squares).  Errors bars correspond to half-sigma error. The dotted (respectively dashed) lines correspond to the mean of the lower (respectively higher) bin mass.
The mean values of  $1+\xi(\cos \theta =0.9)$ are 0.975 for stellar mass $M_{\rm s}=10^{10.75}$ and 1.023  for $M_{\rm s}=10^{10.25}$, respectively.
The transition mass seems reasonably bracketed at $M_{\rm tr, s}=10^{10.5 \pm 0.25} M_\odot$.}
\label{fig:transition-redshift}
\end{figure}

More massive galaxies tend to have their spin preferentially perpendicular to their filament, while less massive ones have their spin preferentially parallel.
A transition occurs around a stellar mass of $M_{\rm tr, s}=3\times 10^{10} \, \rm M_\odot$.
This value is fully consistent with earlier findings of a mass transition for the orientation of the spin of haloes of $M_{\rm tr, h}=5\times 10^{11} \, \rm M_\odot$ at that redshift~\citep{codisetal12} and suggested by the galaxy--halo mass relation determined by abundance matching techniques~\citep{mosteretal13}. 
Using the full redshift sample, Fig.~\ref{fig:transition-redshift} shows that the mass transition appears to be reasonably bracketed at $M_{\rm tr, s}\simeq10^{10.5\pm 0.25} \, \rm M_\odot$.
The mean values of the PDF $1+\xi$ at $\cos \theta =0.9$ are, respectively, 0.98 and 1.02  for  $M_{\rm s}=10^{10.75} \, \rm M_\odot$ and $10^{10.25} \, \rm M_\odot$.

The definition of Hubble type relies on different tracers.
Hence, it is of interest to quantify the alignment or misalignment of galaxies classified according to these tracers.
One should keep in mind that these tracers are not independent from one another (as illustrated in Appendix~\ref{sec:galprops}).
Top row, middle column of Fig.~\ref{fig:fullxi} shows the excess probability of alignment for $V/\sigma$.
Dispersion-dominated galaxies with small $V/\sigma$ ratios (i.e. elliptical galaxies) have their spin perpendicular to filaments, while centrifugally supported galaxies with large $V/\sigma$ (i.e. disc galaxies) have their spins parallel to filaments. 
The transition between parallel and perpendicular alignment occurs at $V/\sigma=0.6$.
A similar signal, not represented here, is found for intrinsic (three-dimensional) kinematics.
The top-right panel of Fig.~\ref{fig:fullxi} shows $\xi$ as a function of the sSFR of galaxies.
Intense star-forming galaxies that rejuvenate their stellar mass content in less than $1/{\rm sSFR}=1/10^{-0.5}\simeq3\,\rm Gyr$ tend to align with filaments.
Conversely, galaxies that are passive (sSFR$\simeq 0.1 \,\rm Gyr^{-1}$) show a random orientation of their spin relative to the filaments.
The left and central panels of the middle row of Fig.~\ref{fig:fullxi} shows $\xi$ as a function of the $g-r$ and the $r-i$ colours of galaxies, respectively.
Redder galaxies ($g-r>0.25$ or $r-i>0.13$) have their spin perpendicular to their filaments, while bluer galaxies ($g-r\le 0.25$ or $r-i\le 0.13$) have their spin parallel to them.
The right-hand panel of the middle row of Fig.~\ref{fig:fullxi} shows $\xi$ as a function of the stellar metallicity $Z$. 
Metal-poor galaxies are more aligned with filaments than metal-rich galaxies which tend to be misaligned.
The bottom-left panel of Fig.~\ref{fig:fullxi} shows $\xi$ as a function of the galaxy age.
Older galaxies have their spin more randomly oriented with that of the filaments, and young galaxies with age below $\simeq 1.2 \, \rm Gyr$ exhibit a stronger alignment.
Finally, the bottom-middle and bottom-right panels of Fig.~\ref{fig:fullxi} show $\xi$ as a function of the $M_{20}$ and Gini quantitative morphological indices.
Galaxies with high $M_{20}$ are more aligned with filaments than galaxies with low $M_{20}$.
Galaxies with low Gini are more aligned with filaments than galaxies with high Gini.
Galaxies with low $M_{20}$ and high Gini tend to trace elliptical galaxies~\citep{Lotzetal08}. 
Note that for age, $M_{20}$ and Gini, galaxies in the \hagn simulation do not seem to present enough leverage to identify a complete 
misalignment (in contrast to the other tracers). 

To summarize, massive, dispersion-dominated, passive, red, smooth, metal-rich and old galaxies tend to have a spin perpendicular or randomly oriented to filaments.
In contrast, low-mass, centrifugally supported, star-forming, blue, irregular, metal-poor and young galaxies tend to align with filaments.

The transitions presented here are more indicative of a trend than a definite proof that each tracer yields a precise morphological transition.
Amongst the various tracers, $V/\sigma$ and $g-r$ are those for which the transition from alignment to perpendicular misalignment is the most significant.
Yet the ensemble allows us to have confidence in the underlying physical picture as they are all consistent with the expected variations.
The above mentioned consistent analysis of its redshift evolution (Fig.~\ref{fig:transition-redshift}) brings further confidence in our results.
It should also be noted that each estimator is derived from a fairly crude automated analysis.

\begin{figure}
\center \includegraphics[width=0.75\columnwidth]{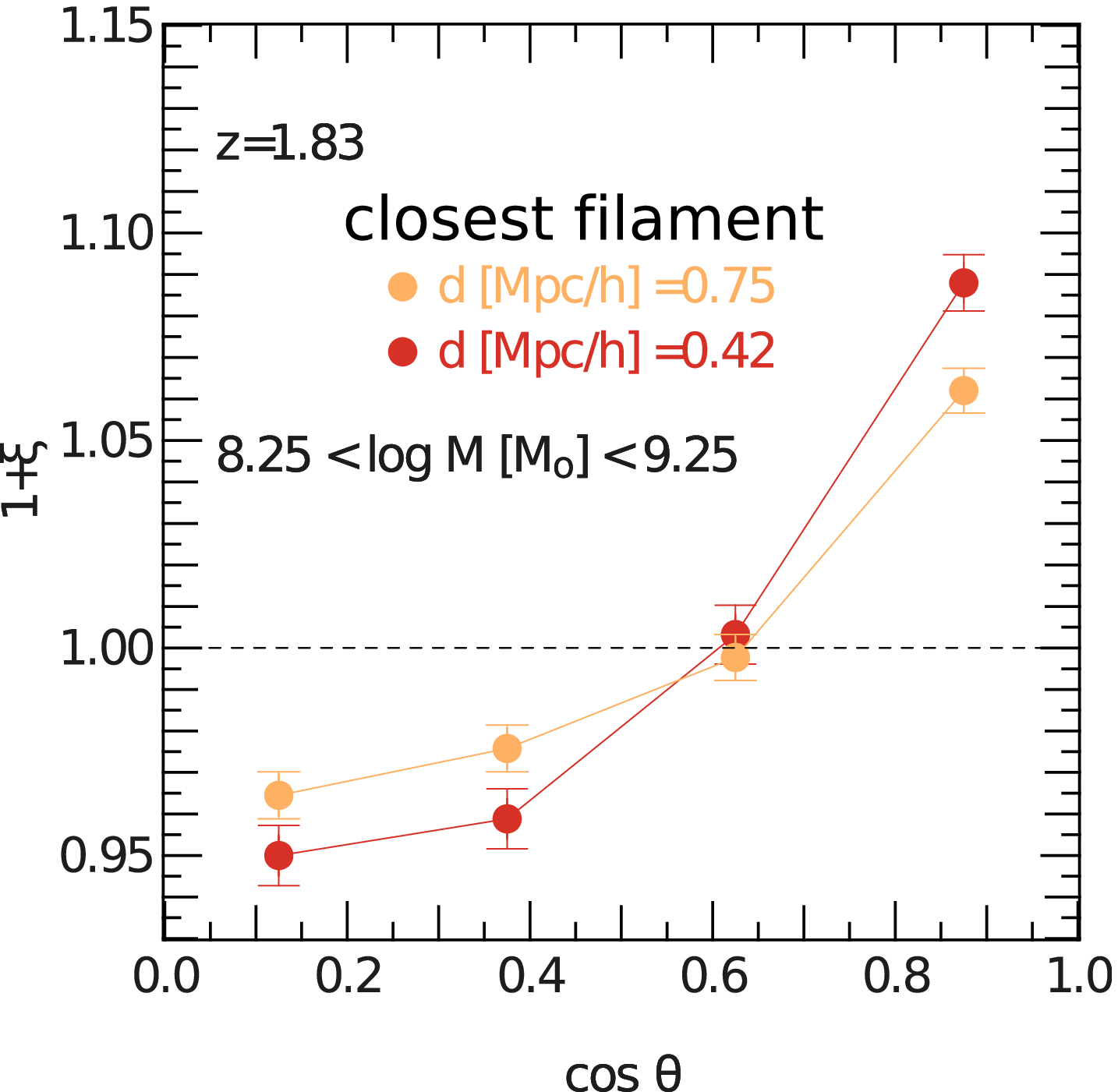}   
  \center \includegraphics[width=0.75\columnwidth]{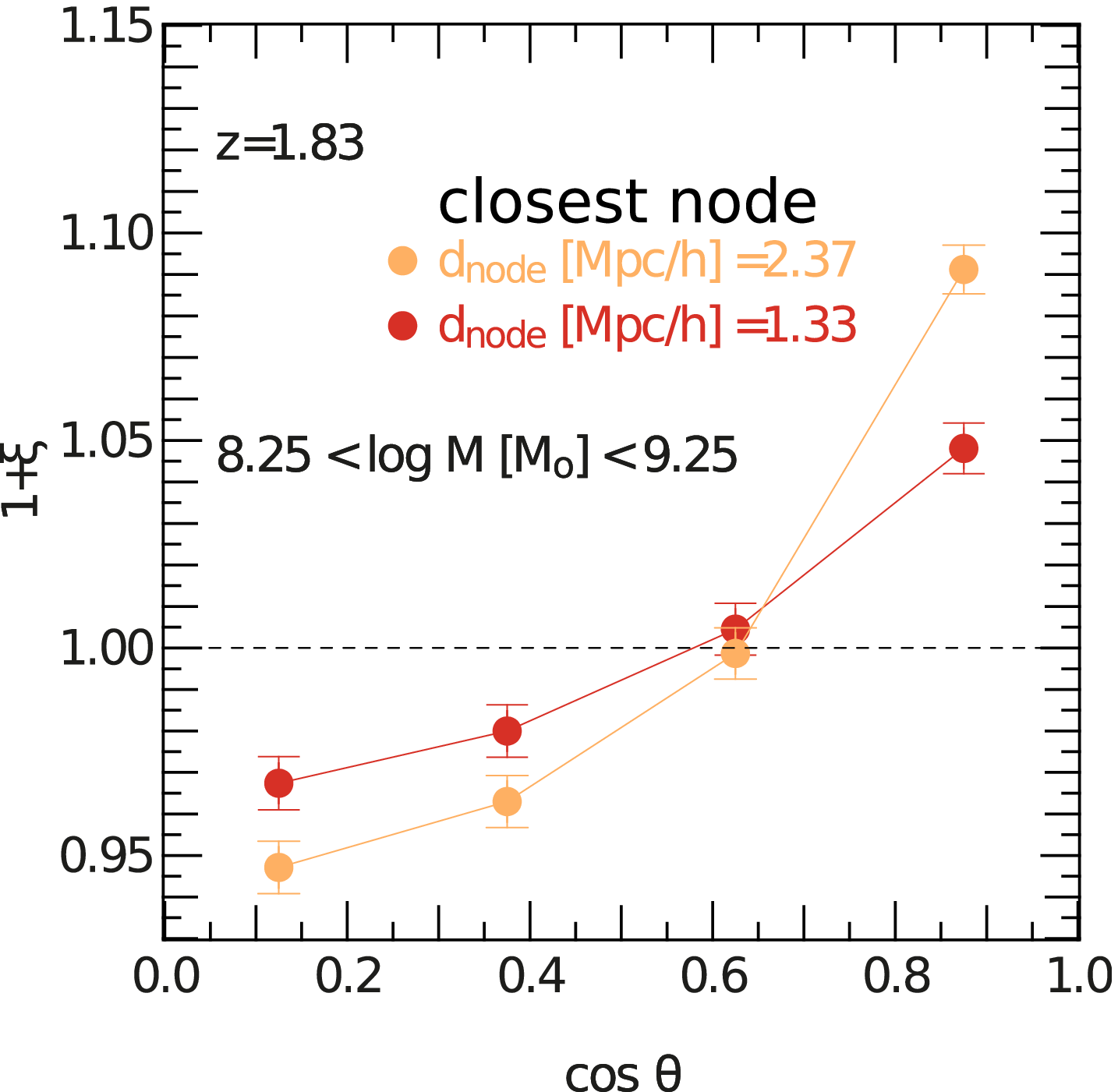}   
        \caption{Excess probability $\xi$ of the alignment between the spin of galaxies and their closest filament as a function of their distance to the closest filament (top panel) or node (bottom panel) is shown at $z=1.83$. Galaxies closer to filaments have their spin more parallel, while galaxies closer to nodes are more randomly oriented. Dashed line is the zero excess probability $\xi =0$.}
\label{fig:dist}
\end{figure}

\subsection{Spin orientation along the cosmic web}
\label{sec:along}

We now investigate the orientation of the alignment as a function of the distance to filaments and nodes.
The upper panel of Fig.~\ref{fig:dist} shows the evolution of alignment of the spin of galaxies as a function of distance to the closest filament for a low-mass subsample.
We apply this measurement to low-mass galaxies because they lie in filaments, sheets and voids, while the most massive galaxies are usually located at the intersection of the most massive filaments in the most massive haloes, therefore, with a minimum scatter in distance to filaments.
Galaxies closer to filaments have their spin more parallel.
The lower panel of Fig.~\ref{fig:dist} shows the evolution of alignment of the spin of galaxies as a function of distance to nodes (i.e. where filaments intersect) along the filaments.
Galaxies further away from nodes have their spin more parallel than galaxies closer to nodes.
This is consistent with the idea that galaxies merge while drifting along filaments (which destroys alignment), and with the strong colour (curvilinear) gradients found by~\cite{gayetal10}.

\begin{figure}
  \includegraphics[width=0.85\columnwidth]{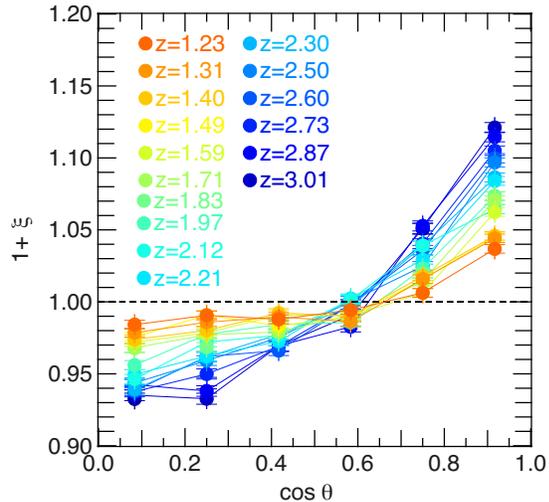}
\caption{Excess probability $\xi$ of the alignment between the spin of galaxies (with mass above $10^8 \, \rm M_\odot$) and their closest filament as a function of redshift is shown. Dashed line is the zero excess probability $\xi =0$. The amplitude of the correlation shows an alignment which increases with redshift (i.e. decreases with cosmic time).}
\label{fig:redshift-evolution}
\end{figure}

\subsection{Redshift evolution}
\label{sec:evolution}

We now investigate the redshift evolution of the excess probability of alignment.
We post-process the \hagn in the following redshift range: $z=3.01-1.23$.
Fig.~\ref{fig:redshift-evolution} shows the amplitude of the alignment of the spin of all galaxies as a function of redshift.
The PDF shows that, on average, galaxies are aligned with their neighbouring filament, because low-mass galaxies dominate in number over massive galaxies (because the mass function of galaxies is strongly decreasing with mass).
The amplitude of the alignment decreases with cosmic time (decreasing redshift) which is a result of more galaxies evolving passively  (i.e. for a given mass, the SFR decreases with time).
The lower the redshift, the stronger the amount of shell crossing and cumulative contribution from mergers along the filaments which tend to destroy the existing alignment and convert orbital momentum into spin to orient it perpendicular to that of the host filament.

\begin{figure}
\includegraphics[width=8cm]{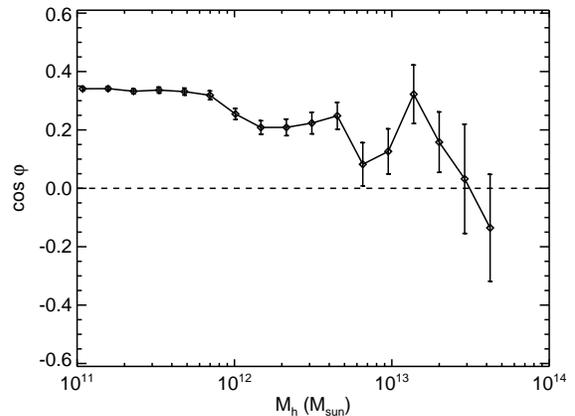}
        \caption{Average cosine of the angle $\cos \psi$ between the galaxy angular momentum and that of its host halo as a function of the halo mass at $z=1.3$ (solid line). Error bars are the standard errors of the mean. Note that the random distribution (dashed line) is for $\cos \psi =0$ because vectors can be pointing towards different directions and, therefore, galaxy spin and halo spin can be aligned or anti-aligned.}
\label{fig:alignhalogal}
\end{figure}

\section{Discussion: AGN feedback promotes spin swings?}
\label{sec:merger}

The above analysis suggests that the most massive galaxies, which are also the reddest, the oldest, the most metal-rich and the most pressure-supported, tend to have their spin perpendicular to the axis of filaments.
Galaxies on average also show less alignment with time (i.e. with decreasing redshift).
We argue that the origin of the misalignment is the sudden reorientation of galactic angular momentum during mergers.
This was shown in~\cite{codisetal12} to be the case for the origin of the DM halo--filament misalignment for massive haloes.
Misalignment occurs because orbital momentum is converted into spin, as DM haloes catch up each other \emph{along} the filaments.
We also argue that indirect merger rate indicators, such as those presented in Section~\ref{section:postprocess} (galaxy properties), can only be modelled once an efficient feedback mechanism is implemented to produce morphological and physical diversity.

\subsection{The contribution of cosmic dynamics}

Massive galaxies undergo mergers (minor and major) that contribute to misalignment of their spin with respect to the direction of the filament.
Given the level of significance for this range of galaxy mass (compatible with a uniform PDF at less than 1-$\sigma$ for $M_{\rm s}\simeq10^{10.75}\, \rm M_\odot$, see the top-left panel of Fig.~\ref{fig:fullxi}), it is still unclear whether massive galaxies have as strong a preference for a spin orientation perpendicular to the axis of their neighbouring filament as do their halo counterparts, or if they tend to be randomly oriented. 

\begin{figure}
\includegraphics[width=8cm]{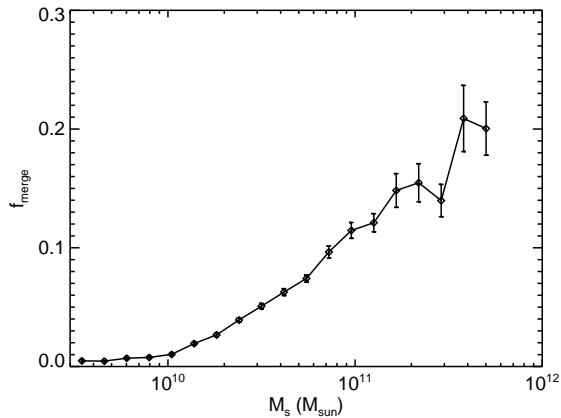}
        \caption{Average fraction of stellar mass gained through mergers as a function of the galaxy stellar mass at $z=1.83$. The error bars are the standard errors on the mean. More massive galaxies have a larger fraction of galaxy mergers contributing to their stellar mass. Lower mass galaxies build up their stellar mass through \emph{in situ} star formation only.}
\label{fig:fmergevsmgal}
\end{figure}

First, note that, globally, the excess probability of alignment is always at values $\xi \lesssim0.1$ (Figs~\ref{fig:fullxi},~\ref{fig:dist} and~\ref{fig:redshift-evolution}), therefore, the signal is weak (not all galaxies align or misalign) but statistically significant.
For instance, the (mis)alignment signal between haloes and filaments~\citep{codisetal12} or that between the large-scale vorticity and filaments~\citep{laigle2014} are respectively of somewhat ($\sim 15$ per cent) and significantly ($\sim 100$ per cent) larger amplitudes.
One could expect that small scales (i.e. galaxies) decouple more strongly from the large-scale filaments than the intermediate scales (i.e. haloes), which are the first virialized structures. 
In particular, there could be a significant amount of redistribution of angular momentum within the inner regions of haloes (\citealp{kimmetal11}; \citealp{Dubois2011}; \citealp{danovichetal11}; \citealp{kassinetal12a}; \citealp{skassinetal12}; \citealp{tillson12}), and as a consequence, 
galaxies misalign with the spin of their host halo (see Fig.~\ref{fig:alignhalogal} for the average cosine of the angle $\psi$ between the angular momentum of the galaxy and that of the host halo), while they keep a strong alignment with the spin of the dark halo's central region~\citep{hahn10}. 
On the other hand, low-mass central galaxies are fed angular momentum directly by cold flows \citep{pichonetal11,tillson12}, which connects them more tightly to their cosmic environment than one would naively expect.
The orientation and amplitude of the stellar component itself reflect the corresponding cumulative advection of  cold gas directly on to the circumgalactic medium.  
In contrast, the orientation of the spin of haloes is more sensitive to the latest stochastic accretion events at the virial radius. 
The net outcome of both competing  processes, as traced by the physical properties of galaxies, is summarized in Figs~\ref{fig:fullxi} and~\ref{fig:transition-redshift}.

\begin{figure}
\includegraphics[width=8cm]{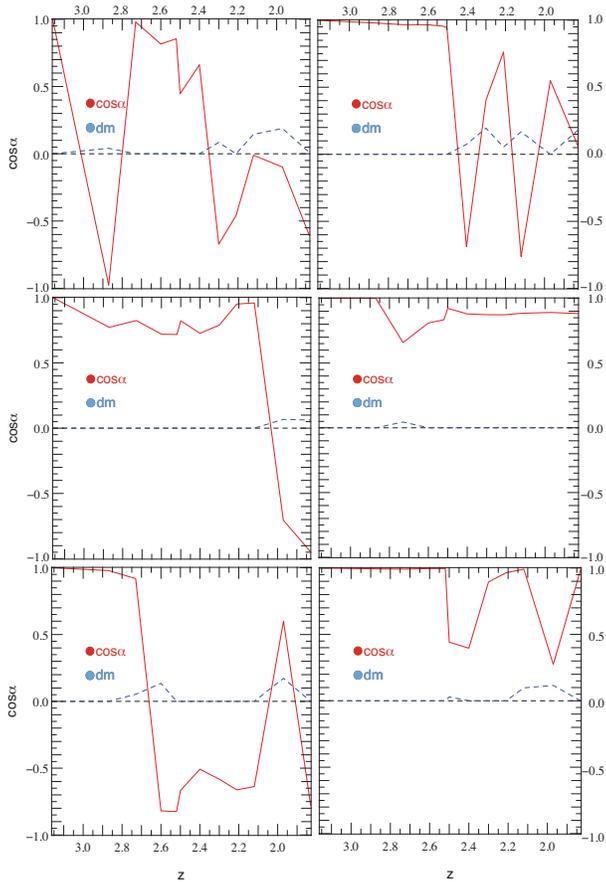}
        \caption{Examples of galaxies changing their spin direction during mergers with stellar mass $1.7\times 10^{11}\, \rm M_\odot$ (top left), $7.3\times 10^{10}\, \rm M_\odot$ (top right), $3.8\times 10^{10}\, \rm M_\odot$ (middle left), $4.8\times 10^{10}\, \rm M_\odot$ (middle right), $1.2\times 10^{11}\, \rm M_\odot$ (bottom left) and $6.0 \times 10^{10}\, \rm M_\odot$ (bottom right) at $z=1.83$. $\cos \alpha$ (red curve) is the cosine of the angle between the spin of the galaxy at the current redshift and the initial spin measured at $z=3.15$. The differential fraction of mass between two time steps coming from mergers ${\rm d}m=\delta M_{\rm merge}/M_{\rm s}$ (in blue) is overplotted. Non-zero values correspond to rapid changes in spin direction. In the absence of mergers, the galaxy spin has a steady direction.}
\label{fig:exemple_spinchange}
\end{figure}

Fig.~\ref{fig:alignhalogal} shows that,
because of the above-mentioned redistribution of angular momentum within the inner region of the halo, the galactic spin is weakly correlated to that of the whole halo, and the effect is more pronounced for more massive haloes which merge more frequently.
Satellites end up reaching the central galaxy with less correlated orbital angular momentum even though they globally originate from a preferred direction, as set by the cosmic web.
In order to test this hypothesis, we build merger trees from the catalogue of galactic structures detected by our galaxy finder.
For each galaxy, we measure the stellar mass acquired through the different branches of the tree (satellites) that we quote as a merger, the main progenitor being excluded from the calculation.
Fig.~\ref{fig:fmergevsmgal} shows that massive galaxies acquire a non-negligible fraction of their mass by mergers 
(at least 1000 particles of star particles, up to 20 per cent at $z=1.83$), while low-mass galaxies grow their stellar mass content almost exclusively by \emph{in situ} star formation~\citep[e.g.][]{delucia&blaizot07, oseretal10}.
Fig.~\ref{fig:exemple_spinchange} shows examples of the evolution of the spin orientation for six massive galaxies, $4\times 10^{10} \lesssim M_{\rm s} \lesssim  2\times 10^{11}\, \rm M_\odot$.
They have a significant contribution from mergers to their stellar mass, which play a significant role in shaping their spin orientations~\citep{bett&frenk12}.
In Fig.~\ref{fig:exemple_spinchange}, the fraction of mass gained by mergers $\delta M_{\rm merge}/M_{\rm s}$ between two time steps is indicated by dashed blue lines.
When no mergers happen, galaxies keep a steady spin direction. 
It is only when a companion galaxy is captured ($\delta M_{\rm merge}/M_{\rm s} \ne 0$) do we see a sudden reorientation of the spin. 
An investigation of the relative role of minor, major, dry and wet mergers is postponed to a companion paper~\citep{welkeretal14} which shows unambiguously that major mergers are indeed responsible for important spin swings.

\begin{figure}
\includegraphics[width=0.9\columnwidth]{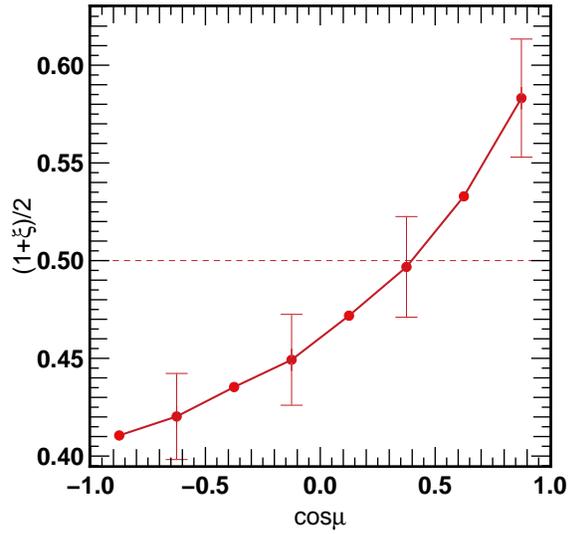} 
\caption{Excess probability $\xi$ of the cosine of the angle $\mu$ between the vorticity of the gas and the direction of the spin of the galaxies at $z=1.83$. The dashed line indicates no correlation. Vorticity is computed as the curl of the velocity field, after a Gaussian smoothing of the velocity field with a kernel length of 780 $h^{-1}$kpc. Note that  fewer galaxies are anti-aligned. Error bars are the standard errors of the mean.}
\label{fig:Vortspin}
\end{figure}

In contrast to high-mass galaxies, low-mass galaxies have their spins preferentially aligned with that of their closest filaments. 
Gas embedded within large-scale walls streams into the filaments which bound them, winding up to form the first generation of galaxies which have spins parallel to these filaments~\citep{pichonetal11}.
Since these galaxies build up their stellar mass \emph{in situ} without significant external perturbations, the stars retain the angular momentum of the cold gas obtained directly from the cosmic web. 
Fig.~\ref{fig:Vortspin} shows the excess probability of the cosine of the angle $\mu$ between the vorticity of the gas (as estimated on scales of $200 \, h^{-1}\, \rm kpc$) at the galaxy's position and the direction of the spin of galaxies (dominated by the low-mass population).
As was found in \cite{laigle2014} for the spin of DM haloes (see also~\citealp{libeskind13}), the galactic spin is also strongly correlated with the vorticity of the surrounding gas.
This correlation has polarity: there are fewer galaxies with their spin anti-aligned with the local vorticity.  
This  dynamical and stellar evidence therefore allows us to apply to baryons the scenario presented in~\cite{laigle2014} on the vorticity-driven origin of the galactic spin--filament alignment.

\subsection{The contribution of AGN feedback}

\begin{figure}
\includegraphics[width=8cm]{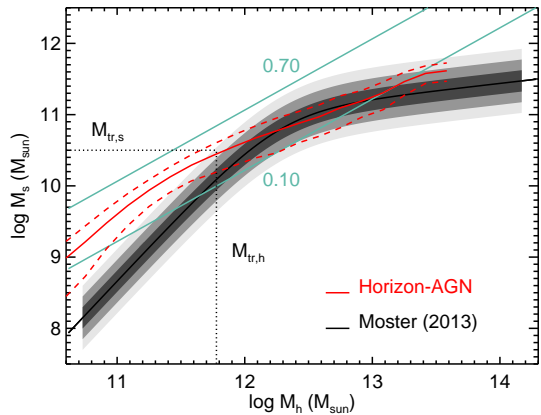}
        \caption{Average stellar mass of the central galaxy as a function of the host halo mass in the simulation at $z=1.3$ (red curve) with the $1\sigma$ dispersion together with the abundance matching result from~\citet{mosteretal13} at the same redshift assuming a Salpeter IMF ($+0.19 \, \rm dex$ in stellar mass from Kroupa to Salpeter IMF) with the $1\sigma$, $2\sigma$ and $3\sigma$ error bars represented by the shaded areas. The cyan lines indicate constant stellar efficiencies defined as $f_{\rm eff}=M_{\rm s}/(\Omega_{\rm b}/\Omega_{\rm m} M_{\rm h})$. The transition mass between alignment and misalignment of the galaxy ($M_{\rm tr,s}$), respectively, halo ($M_{\rm tr,h}$), spin and the filament are represented as dotted lines. }
\label{fig:mstarvsmhalo}
\end{figure}

\begin{figure}
\includegraphics[width=8cm]{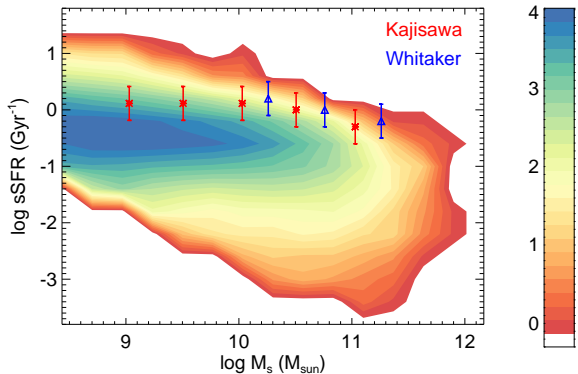}
        \caption{Diagram of the distribution of galaxies as a function of their sSFR and stellar mass $M_{\rm s}$ at $z=1.3$. The contours are the number of galaxies in log units. Red and blue points with error bars are the observations at $z=1.25$ extracted from~\citet{behroozietal13} and originally from \citet{kajisawaetal10} and \citet{whitakeretal12} and rescaled for a Salpeter IMF ($+0.26 \, \rm dex$ in stellar mass from Chabrier to Salpeter IMF). The sSFR decreases with galaxy stellar mass.}
\label{fig:ssfrvsmgal}
\end{figure}

Fig.~\ref{fig:mstarvsmhalo} shows the stellar halo mass relation at $z=1.3$ for the \hagn simulation.
It appears that above $M_{\rm h}\gtrsim 10^{12}\,\rm M_\odot$ the relation is in good agreement with abundance-matching results from~\cite{mosteretal13}.
Note that the stellar masses from~\cite{mosteretal13} are rescaled by $+0.19 \, \rm dex$ to account for a change from Kroupa to Salpeter IMF.
The presence of the feedback from AGN reduces the amount of stars formed in these massive galaxies, allowing them to agree with observations by reducing the stellar mass up to one order of magnitude~\citep{Dubois2013, puchwein&springel13, martizzietal14}.
Note also that the transition mass from alignment to misalignment is close to the mass ($\simeq 10^{11} M_\odot$) where passive galaxies become more abundant than star-forming galaxies in observations~\citep{Droryetal09, Davidzonetal13, Ilbertetal13}.
For massive galaxies, mergers are key in both swinging the spin \emph{and} changing the morphology. 
The sSFR decreases significantly  with galaxy stellar mass (see Fig.~\ref{fig:ssfrvsmgal}) because of the quenching of their gas accretion rate -- in the shock-heated mode of accretion for massive haloes~\citep[e.g.][]{birnboim&dekel03, ocvirketal08, dekeletal09} -- and because of the strong suppression of cold gas within galaxies via AGN feedback.
Observational data from~\cite{behroozietal13} at $z=1.3$ (originally from \citealp{kajisawaetal10} and \citealp{whitakeretal12}) are represented on top of the distribution of our simulations points.
Stellar masses from~\cite{behroozietal13} are rescaled by $+0.26 \, \rm dex$ to account for a change from Chabrier to Salpeter IMF. 
The quenching of the SFR in massive galaxies through AGN feedback leads to an enhanced fraction of stars gained through mergers~\citep{Dubois2013}.
For lower mass galaxies, the agreement of the stellar halo mass relation with observational data is less favourable because feedback from SNe is not strong enough to suppress the star formation in dwarfs.
Some missing physical processes, such as radiation from young stars, are probably necessary to further suppress the star formation in low-mass galaxies~\citep[e.g.][]{hopkinsetal11, murrayetal11, hopkinsetal13}.
~\cite{vogelsbergeretal13} manage to reproduce the low-mass tail of the stellar-to-halo mass relation by decoupling hydrodynamically the launched wind gas from the dense star-forming gas (as introduced by~\citealp{springel&hernquist03}).
This decoupling of gas is known to generate a more efficient transfer of energy from SNe to large-scale galactic winds, compared to local prescriptions as we have adopted here (as shown by~\citealp{dallavecchia&schaye08}), but it lacks physical motivation.
Note that the choice of the Salpeter IMF instead of a Chabrier IMF decreases the total energy released by a stellar particle by a factor of 3 (assuming a minimum and maximum mass of $0.1\,Ê\rm M_\odot$ and $100\,Ê\rm  M_\odot$).
However, we are still confident that stronger feedback in low-mass galaxies should not drastically change their orientation, since mergers for that class of haloes are rare enough~\citep{fakhourietal10, geneletal10}. This is  in particular true for the major mergers that are required to significantly reorient  the spins of galaxies~\citep{welkeretal14}.

The significant contribution from AGN feedback at reducing the stellar activity in massive galaxies is mandatory to obtain a diversity in the physical properties of galaxies (colours, $V/\sigma$, sSFR, etc.) across the whole mass range (see Appendix~\ref{sec:galprops}).
The effect of AGN feedback is to also reduce the mass of stars formed \emph{in situ}, i.e. to prevent the formation of a rotation-supported component in massive galaxies and to turn massive galaxies into pressure-supported ellipticals~\citep{Dubois2013}.
In the absence of a central supermassive BH, the magnitude 
of the angular momentum of the stellar component of massive galaxies will therefore be larger, 
as a larger fraction of their (larger) stellar mass will be distributed in a rotationally supported disc.
Thus, a merging satellite  the angular momentum of which   is misaligned with that of the central galaxy produces a variation in the angle between the pre-merger and the post-merger spin of the galaxy that is smaller for the disc case (no AGN case) than for the elliptical case (AGN case).
Moreover, for massive galaxies, the feedback from the central AGN switches off later accretion of circumgalactic gas~\citep{duboisetal10}.
Consequently, the possible realignment of the galactic spin with the filament that could operate after a merger due the accretion of fresh gas  is 
 reduced by the presence of the AGN feedback from the galaxy remnant.
AGN feedback thereby acts as a catalyst of spin swings.
  
\section{Conclusions}
\label{sec:conclusion}

Our analysis of the \hagn flagship simulation, which models AGN as well as stellar feedback so as to produce morphological diversity, shows that the orientation of the spin of galaxies depends on various galaxy properties such as stellar mass, $V/\sigma$, sSFR, $M_{20}$, metallicity, colour and age.
The spins of galaxies tend to be preferentially parallel to their neighbouring filaments, for low-mass, young, centrifugally supported, metal-poor, bluer galaxies, and perpendicular for higher mass,  higher velocity dispersion, red, metal-rich old galaxies.
The alignment is the strongest, the closer to the filaments and further from the nodes of the cosmic web the galaxies are.
This is in agreement with the predictions of~\cite{codisetal12} for DM haloes.
We find a transition mass, $M_{\rm tr, s} \simeq 3\times 10^{10}\, \rm M_\odot$ which is also consistent with these authors' predictions for the corresponding halo transition mass.
Due to the weak galaxy--halo alignment, the amplitude of the correlation with cosmic filaments is somewhat weaker for galaxies than for haloes.
It also decreases with cosmic time due to mergers and quenching of cold flows and star formation.
Hence, our results suggest that galaxy properties can be used to trace the spin swings along the cosmic web.

The transition from the aligned to the misaligned case is dynamically triggered by mergers (the frequency of which increases with galaxy mass) that swing the spin of galaxies.
AGN feedback has a key role at preventing further gas inflow and quenching the \emph{in situ} star formation after such an event, in order to preserve the misalignment operated by the merger.

Though it is expected that galaxy modelling will improve over the next decade -- in particular the way feedback is taken into account in large-scale cosmological simulations --  we do not anticipate that the particular results found  in this paper should  change qualitatively.
The finding that the morphological diversity of galaxies traces  populations with different spin--filament alignments, which in turn is in part inherited from the anisotropy of the embedding cosmic web, is both robust predictions of the current gravitational clustering scenario and of our understanding of  the dynamics of elliptical and spiral galaxies. 

In a forthcoming paper, we will analyse more realistic mock catalogues from the \hagn light-cone down to a lower redshift to investigate the amount of modification of the signal induced by dust extinction, projection effects, limited resolution and finite signal-to-noise ratio. 
More efficient and robust estimators for morphology, either intrinsic using the full data set of the simulation or projected using virtual degraded observables, will be built and compared.
Quantitative comparisons to observations are postponed to this paper.

\section*{Acknowledgements}
We thank the anonymous referee for suggestions, which improved the clarity of the paper.
This work has made use  of the HPC resources of CINES (Jade supercomputer) under the allocation 2013047012 made by GENCI.
The post-processing made use of the {\tt horizon} and {\tt Dirac} clusters.
This work is partially supported by the Spin(e) grants ANR-13-BS05-0002 of the French \emph{Agence Nationale de la Recherche} and by the National Science Foundation under Grant No. NSF PHY11-25915.
This research is part of the Horizon-UK project.
YD and CP thank the KITP for hospitality when this project was initiated. 
CP also thanks the institute of Astronomy for a Sacker visiting fellowship and the PEPS `Physique th\'eorique et ses interfaces' for funding.
The research of YD and JS has  been supported at IAP by  ERC project 267117 (DARK) hosted by Universit\'e Pierre et Marie Curie - Paris 6.
YD  is grateful to the Beecroft Institute for Particle Astrophysics and Cosmology  for hospitality  and to the Balzan Foundation for  support, at Oxford University  where some of  this work was carried out.  
The research of JS has also been supported  at the Johns Hopkins University by National Science Foundation grant OIA-1124403.
The research of AS and JD at Oxford is supported by the Oxford Martin School and Adrian Beecroft. 
This work was also supported by the ILP LABEX (under reference ANR-10-LABX-63) was supported by French state funds managed by the ANR within the Investissements d'Avenir programme under reference ANR-11-IDEX-0004-02.
We thank D.~Munro for freely distributing his {\sc \small Yorick} programming language and opengl interface (available at {\tt http://yorick.sourceforge.net/}).

\bibliographystyle{mn2e}
\bibliography{author}

\vspace*{6pt}
\noindent$^{1}$ Sorbonne Universit\'es, UPMC Univ Paris 06, UMR 7095, Institut d'Astrophysique de Paris, F-75005 Paris, France\\
$^{2}$ CNRS, UMR 7095, Institut d'Astrophysique de Paris, 98 bis Boulevard Arago, F-75014 Paris, France\\
$^{3}$ Sub-department of Astrophysics, University of Oxford, Keble Road, Oxford OX1 3RH, UK\\
$^{4}$ KITP Kohn Hall, 4030 University of California Santa Barbara, CA 93106-4030, USA\\
$^{5}$ Institute of Astronomy and Kavli Institute for Cosmology, Madingley Road, Cambridge CB3 0HA, UK\\
$^{6}$ Observatoire de Lyon, UMR 5574, 9 avenue Charles Andr\'e, F-69561 Saint Genis Laval , France\\
$^{7}$ Department of Physics, University of Alberta, 11322-89 Avenue, Edmon ton, Alberta T6G 2G7, Canada\\
$^{8}$ Aix Marseille Universit\'e, CNRS, Laboratoire d'Astrophysique de Marseille, UMR 7326, 38 rue F. Joliot-Curie, 13388 Marseille, France\\
$^{9}$ D\'epartement de Physique Th\'eorique, Universit\'e de Gen\`eve 24 quai Ernest Ansermet, CH-1211 Gen{\`e}ve, Switzerland\\
$^{10}$ Space Telescope Science Institute, 3700 San Martin Drive, Baltimore, MD 21218, USA\\
$^{11}$ Department of Astrophysical Sciences, Princeton University, Peyton Hall, Princeton, NJ 08544, USA\\
$^{12}$ Canada--France--Hawaii Telescope Corporation, 65-1238 Mamalahoa Hwy, Kamuela, HI 96743, USA\\
$^{13}$ Department of Physics and Astronomy, The Johns Hopkins University Homewood Campus, Baltimore, MD 21218, USA\\
$^{14}$ Institute f\"ur Theoretische Physik, Universit\"at Z\"urich, Winterthurerstrasse 190, CH-8057 Z\"urich, Switzerland\\
$^{15}$ Astronomy Department, University of Michigan, Ann Arbor, MI 48109, USA\\

\appendix

\section{Properties of galaxies}
\label{sec:galprops}

\begin{figure*}
\includegraphics[width=0.66\columnwidth]{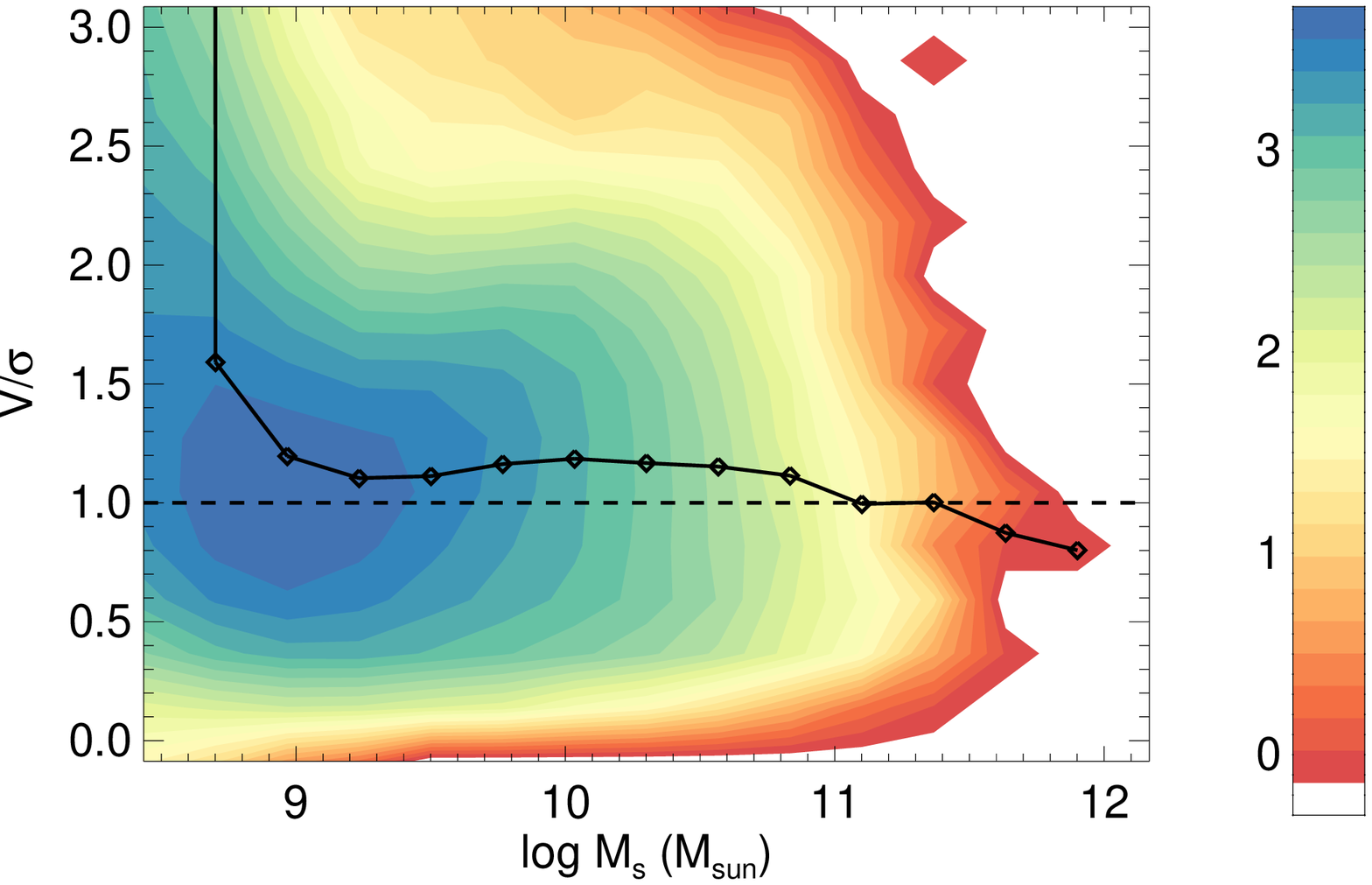}
 \includegraphics[width=0.66\columnwidth]{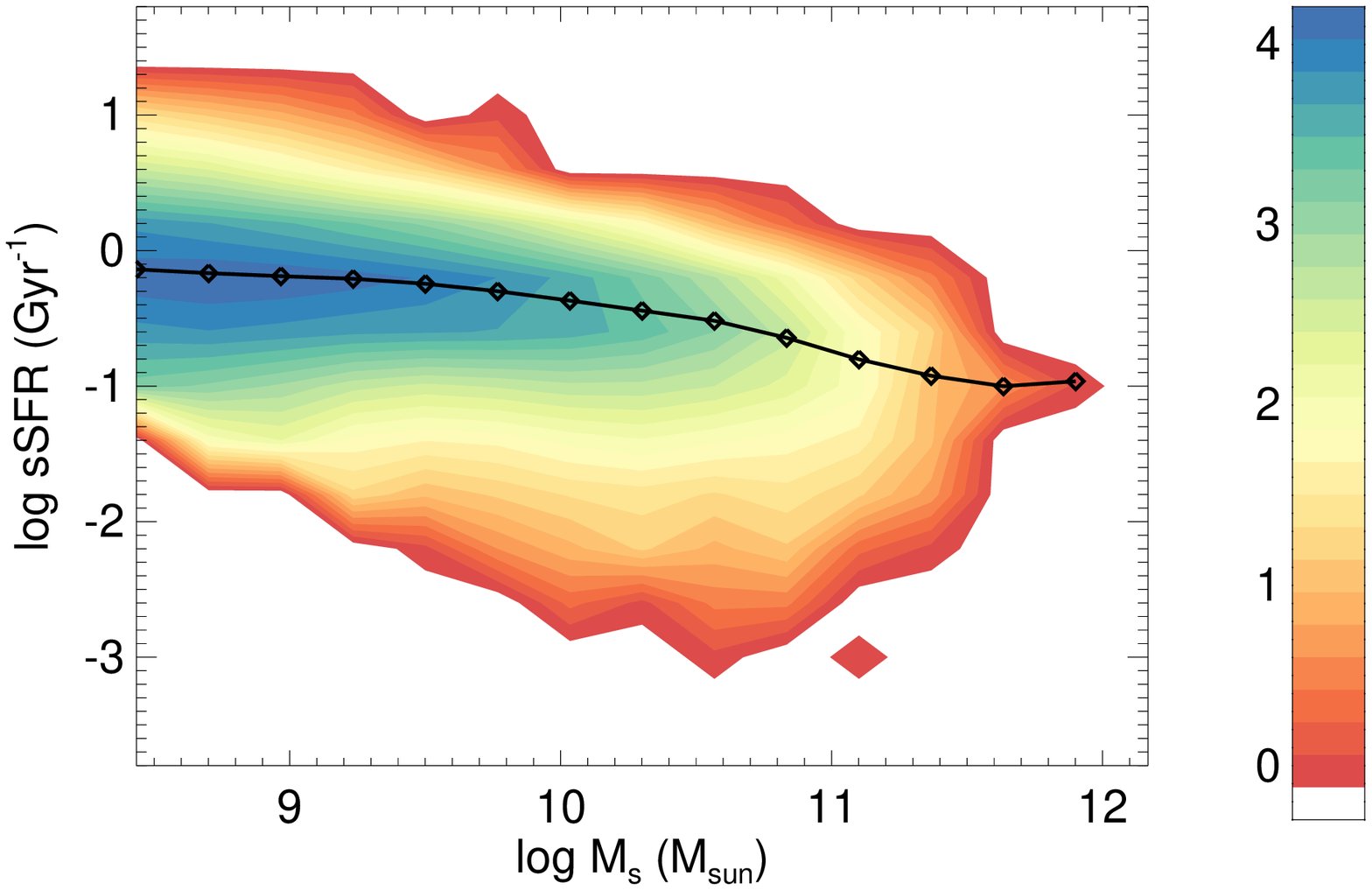}
 \includegraphics[width=0.66\columnwidth]{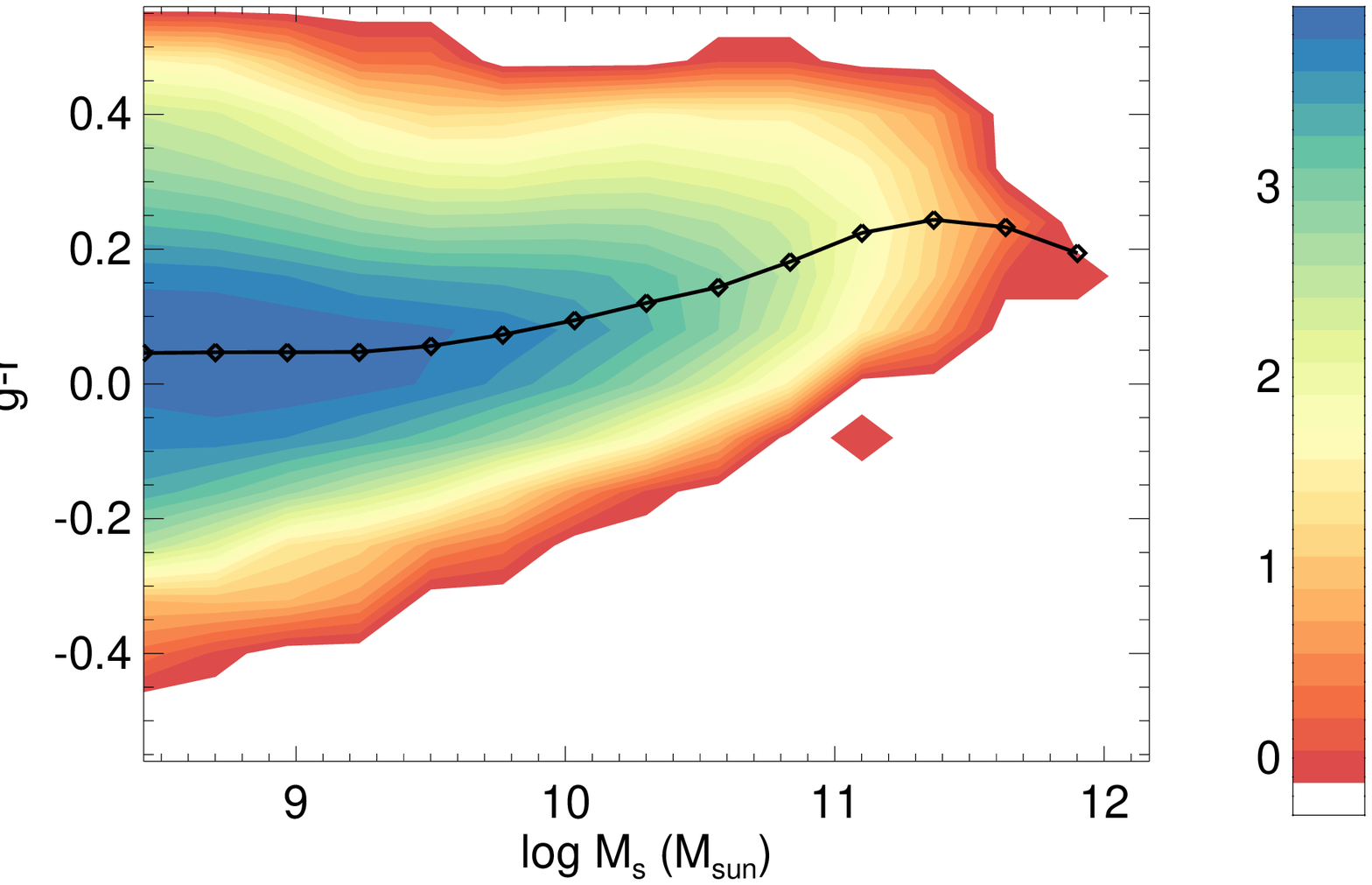}\vspace{-0.4cm}
 \includegraphics[width=0.66\columnwidth]{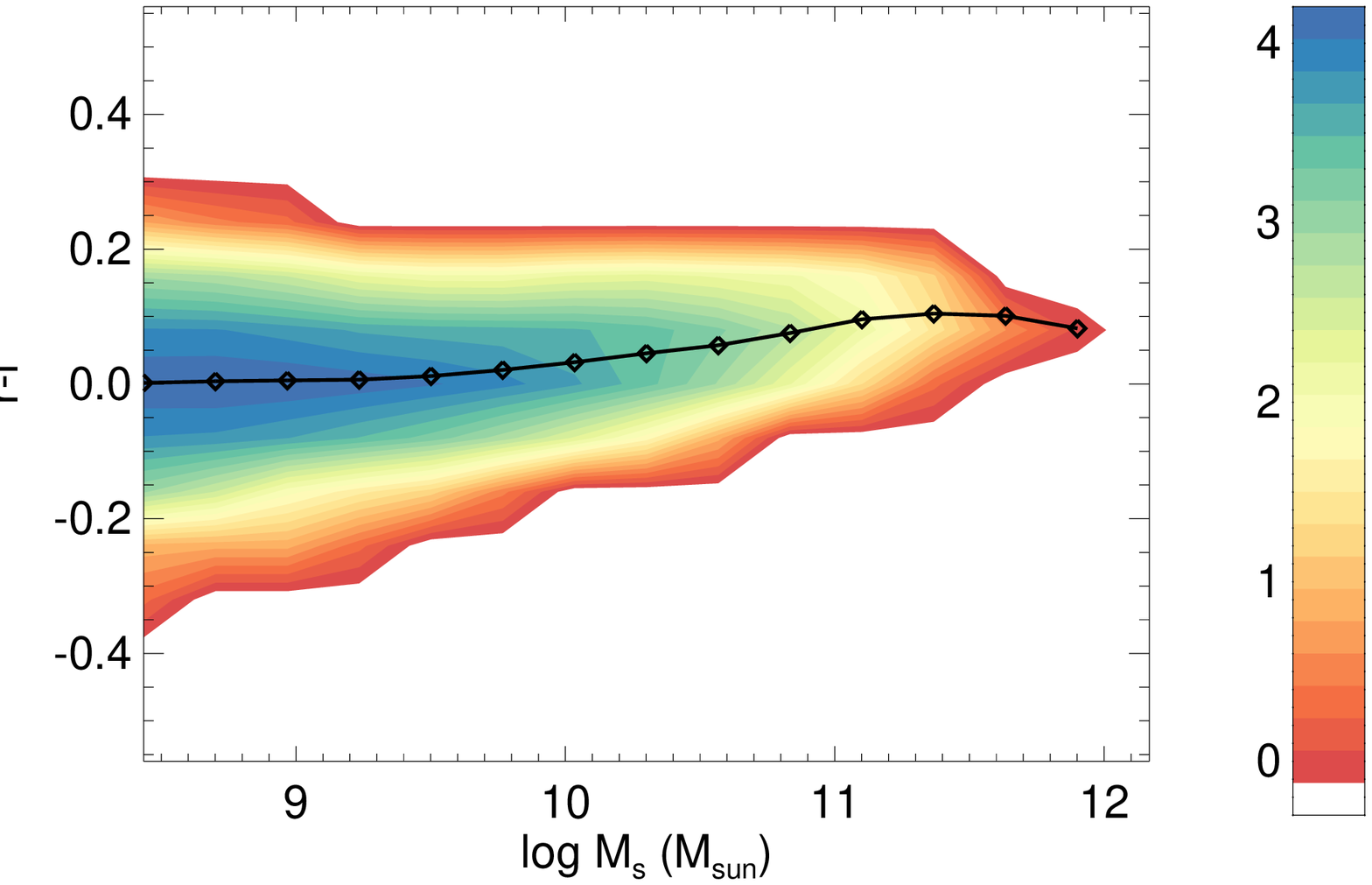}
\includegraphics[width=0.66\columnwidth]{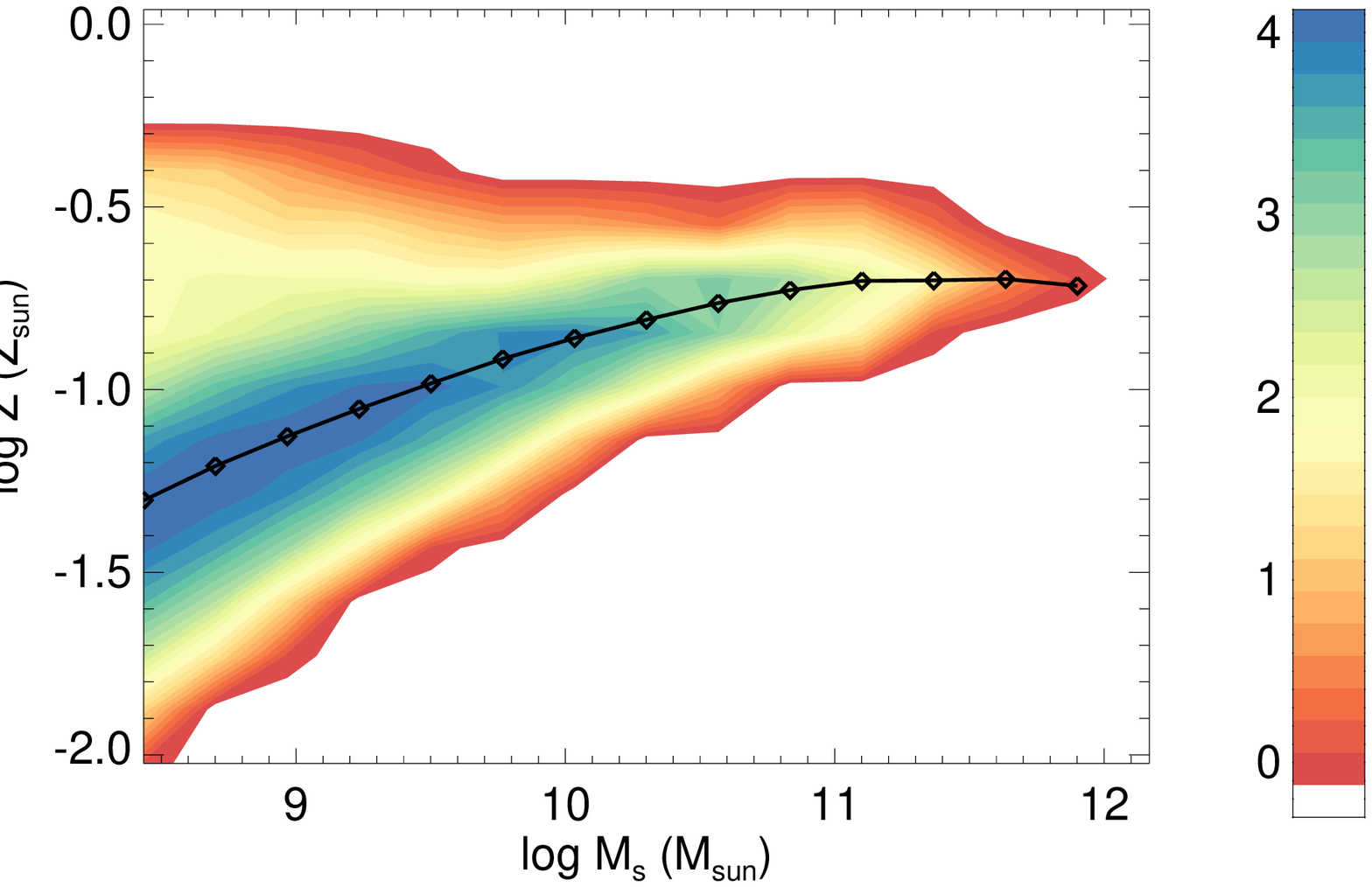}
 \includegraphics[width=0.66\columnwidth]{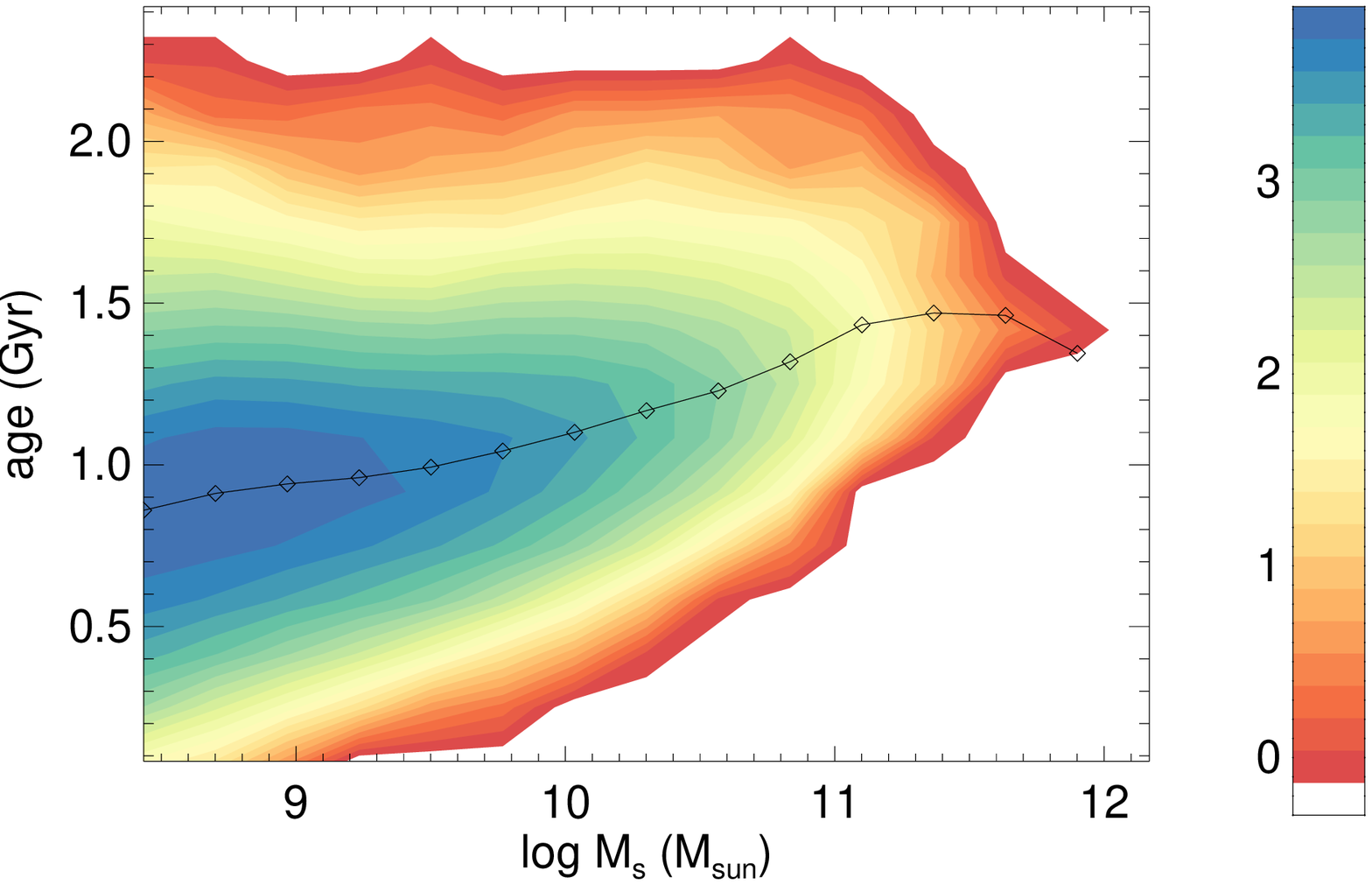}
\caption{From top to bottom and left to right: Contours of the logarithmic number of galaxies for 2D projected kinematics of the stars $V/\sigma$, sSFR, $g-r$ colour, $r-i$ colour, metallicity $Z$ and age as a function of the stellar mass of the galaxy at $z=1.83$. The solid lines with diamonds correspond to the average values as a function of the stellar mass. The dashed line in the top-left panel corresponds to $V/\sigma=1$.}
\label{fig:galprops}
\end{figure*}

Fig.~\ref{fig:galprops} shows various physical properties of galaxies: $V/\sigma$, sSFR, $g-r$ colour and age as a function of the stellar mass at $z=1.83$.
There are correlations between the stellar kinematics $V/\sigma$ and galaxy masses: more massive galaxies are pressure supported; their sSFR and mass: more massive galaxies have lower sSFR, i.e. are more passive; their colour and mass: more massive galaxies are redder; and their age and mass: more massive galaxies are older.
Note the quite large scatter at all masses, that is more pronounced for lower mass objects, which can be explained by the fact that low-mass field galaxies in low-density environments evolve differently from low-mass satellites galaxies in high-density environments.

\label{sec:morpho}

\begin{figure}
  \includegraphics[width=\columnwidth]{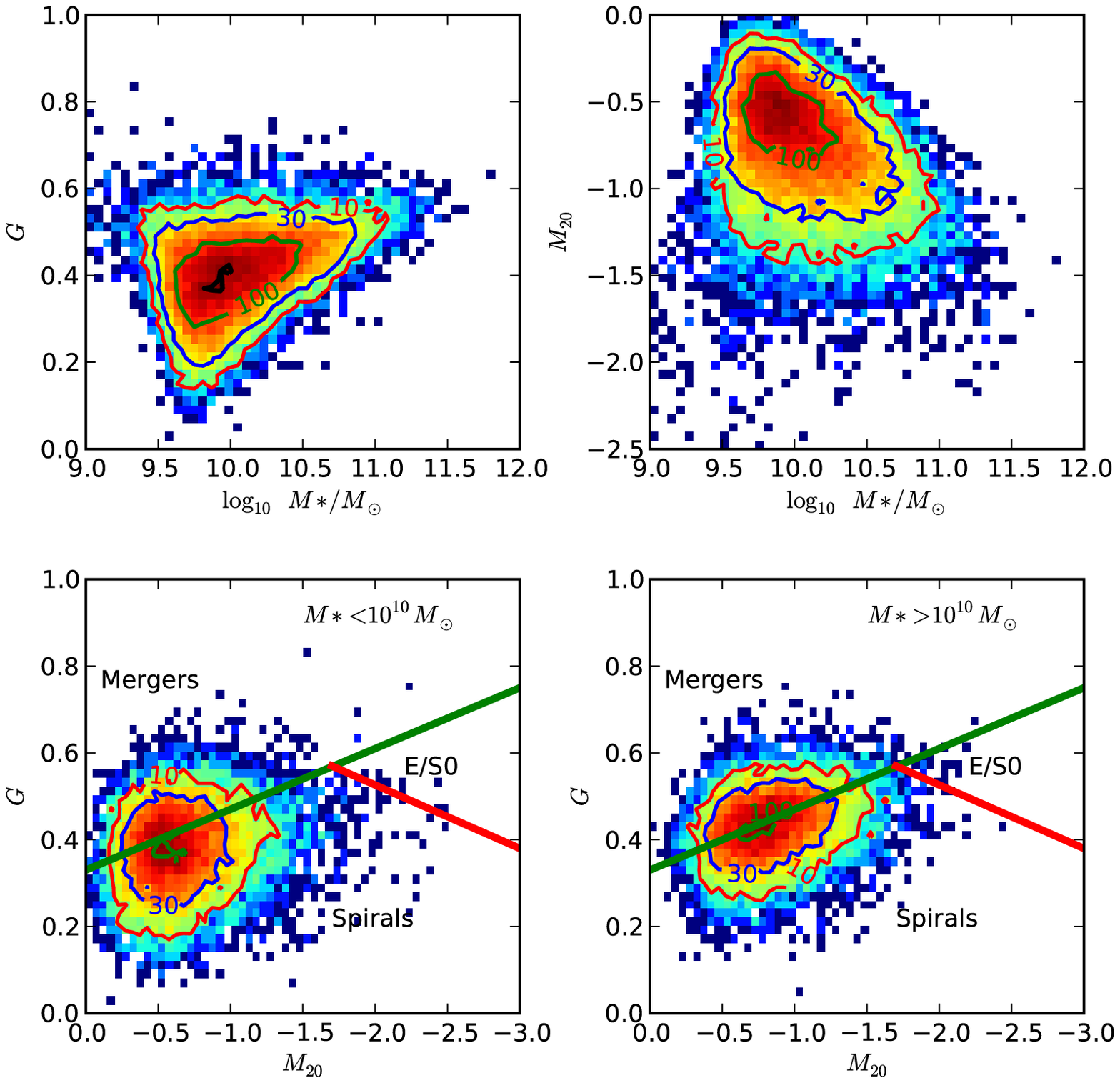}
\caption{Bivariate distributions of stellar mass, Gini ($G$) and $M_{20}$ morphological parameters for galaxies in the simulation at $z=1.8$ in the rest-frame $u$ filter. Lower panels show the Gini--$M_{20}$ distributions for stellar mass selected samples. The zones for elliptical, spiral and merger galaxies are schematically drawn as tentative locations of local ellipticals, spirals and mergers at $z\simeq2$. See the text for details.}
\label{fig:GiniM20mass}
\end{figure}

Fig.~\ref{fig:GiniM20mass} shows the relation between $M_{\rm s}$ and the morphology parameters Gini and $M_{20}$ measured in the rest-frame $u$ band. These morphological estimators allow in principle to separate spirals from spheroids, ellipticals, and merging galaxies. This has been done in the local universe \citep{lotzetal04} and at higher redshift \citep{abrahametal07, Lotzetal08, leeetal13}. The regions drawn in the figure are taken from \citet{Lotzetal08}. Their precise locations with respect to the distributions measured  here should be taken with caution: extinction by dust is not taken into account, morphological $k$-corrections (although quite small) are not accounted for and the spatial resolution is not matched. Despite all these caveats which may explain the relatively small number of galaxies classified as ellipticals or spheroids from these diagrams, there seems to be, qualitatively, rather good agreement between the distributions measured in the deep surveys and in the simulation. 

\section{Grid-locking}
\label{sec:spinlock}

\begin{figure}
\center \includegraphics[width=\columnwidth]{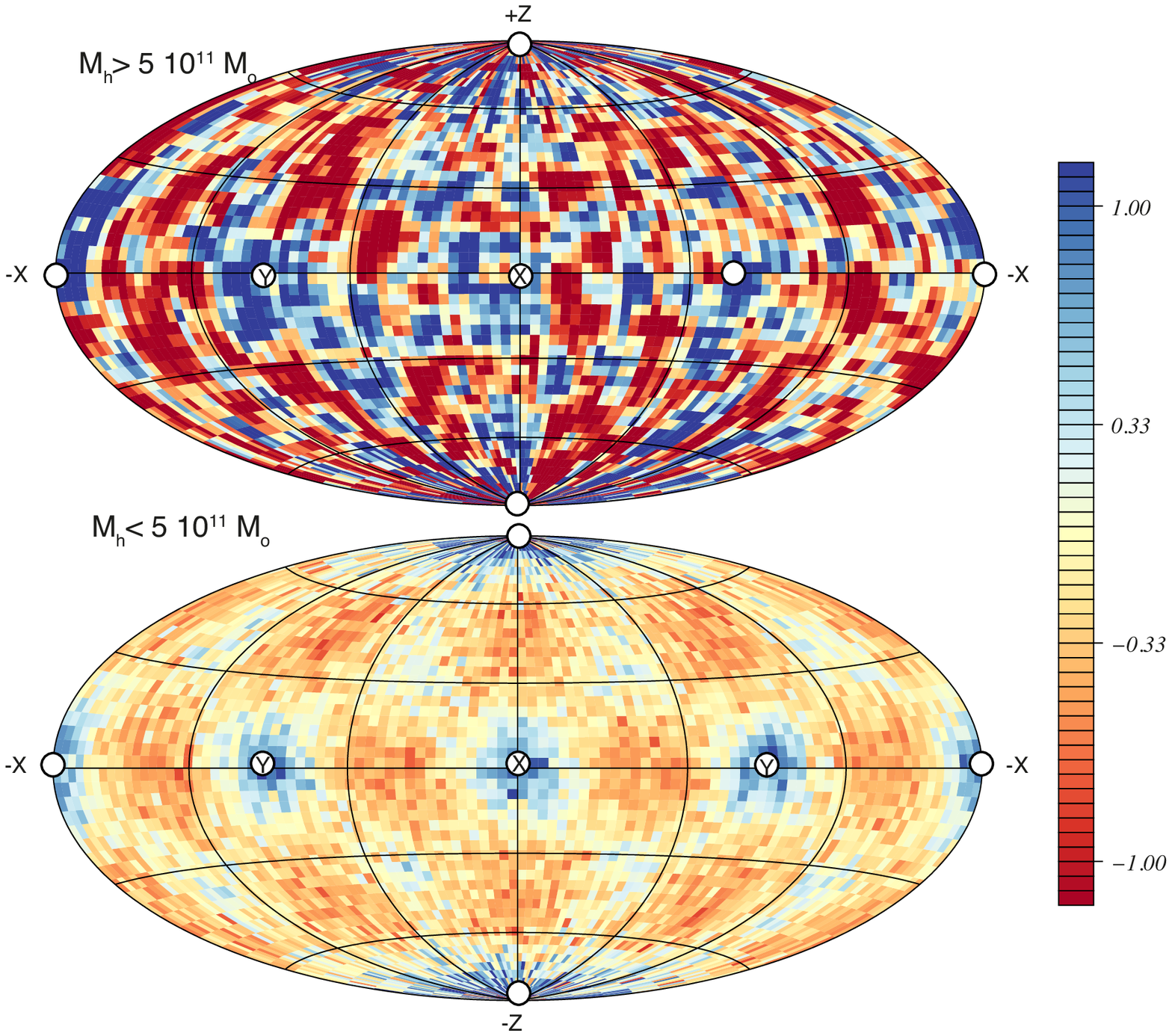}
\caption{The effect of grid-locking at redshift $z=1.3$ of the spin of  \emph{galaxies} for haloes more (top panel), respectively less (bottom panel) massive than $5\times 10^{11}\, \rm  M_\odot$ respectively. The white discs represent the directions of the simulation box as labelled. The colour coding represents relative fluctuations around the mean. The smaller galaxies (bottom panel) show a clear sign of grid-locking, while the more massive sample (top panel) does not.}
\label{fig:mollview}
\end{figure}

Fig.~\ref{fig:mollview} displays a Mollweide projection of the orientations of \emph{galaxy} spins along Cartesian axes, for a range of halo mass. 
Galaxies hosted by haloes lighter or heavier than  $5\times\, \rm  10^{11} M_\odot$ are considered.
While the spins of the less massive galaxies are clearly aligned with the grid, no obvious alignment is seen for the high-mass galaxies.
Lighter galaxies are preferentially locked with the grid because they are composed of very few grid elements:
the gaseous disc of a galaxy with $\sim10^9\, \rm M_\odot$,  embedded in a halo of mass $\sim 10^{11}\,Ê\rm M_\odot$, tends to be aligned with one of the Cartesian axes due to the anisotropic numerical errors. 
However, for more massive galaxies, the grid-locking is absent due to a larger number of resolution elements to describe those objects.
This result is consistent with that of~\cite{hahn10} and~\cite{danovichetal11}.

\begin{figure}
\center \includegraphics[width=0.9\columnwidth]{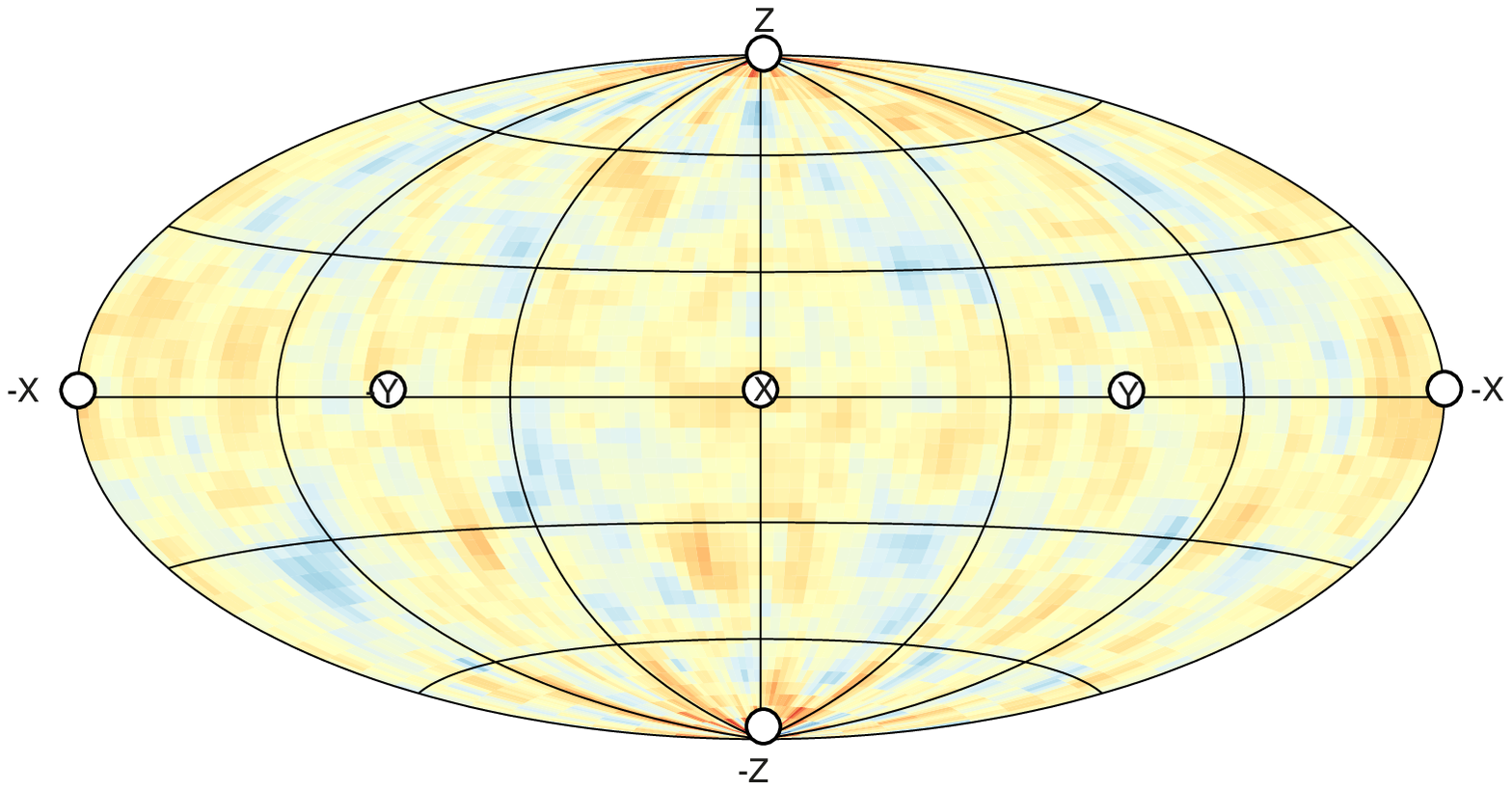}
\caption{The effect of grid-locking of the skeleton segments at $z=1.3$. The colour coding is the same as in Fig.~\ref{fig:mollview}. No preferred direction is shown for the skeleton.}
\label{fig:mollview-skel}
\end{figure}

Fig.~\ref{fig:mollview-skel} shows the distribution of the axis of all the elements of the skeleton on the sphere.
There is no preferential direction of alignment with respect to the  box axes.

\begin{figure}
\center \includegraphics[width=6cm]{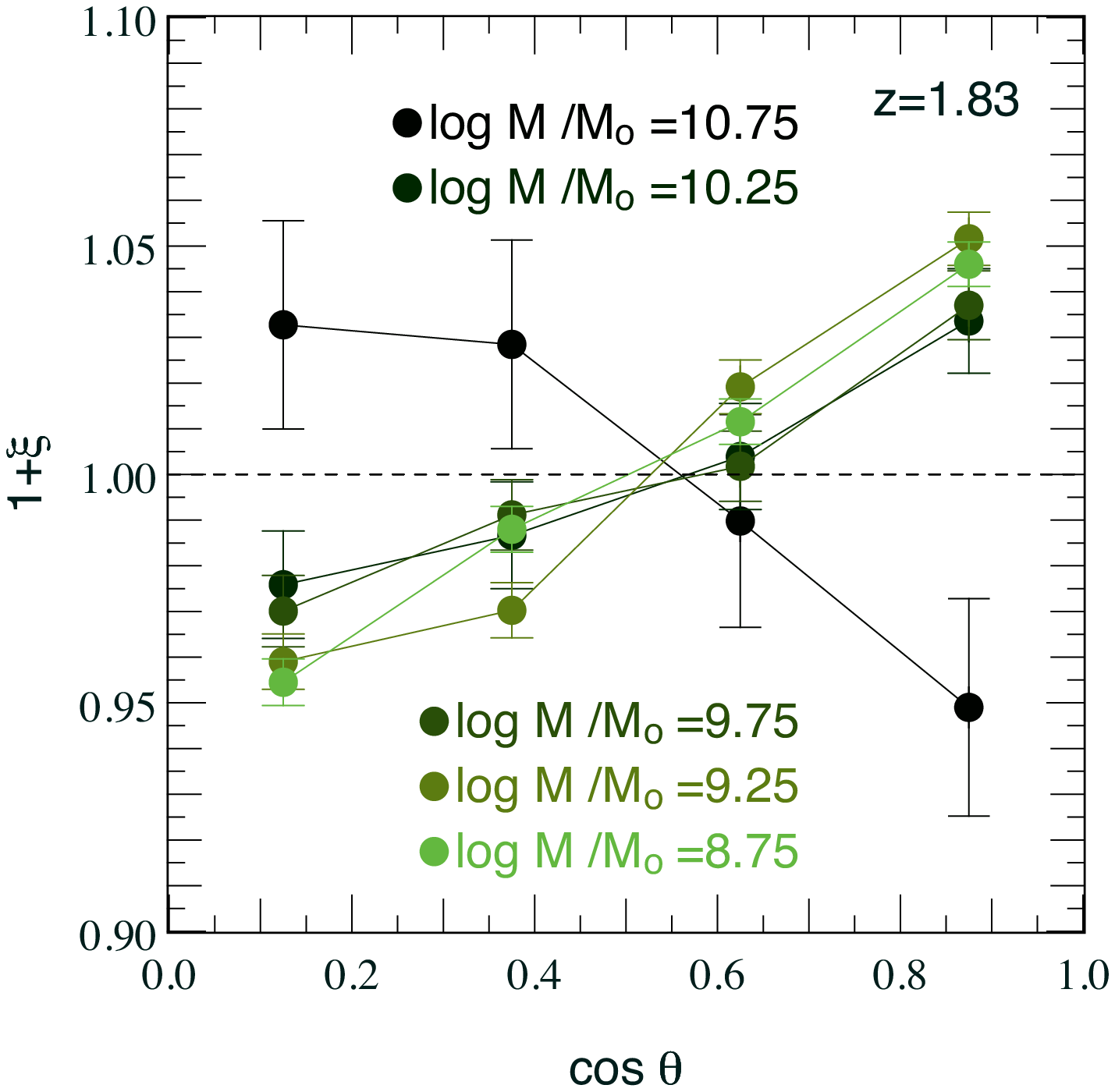}
\caption{Excess probability $\xi$ of the alignment between the spin of galaxies and their filament for galaxies with different stellar masses. Galaxies whose spin is contained within an angle smaller than 10 degrees from any cartesian planes of the box are not taken into account. Dashed line is the zero excess probability $\xi =0$.}
\label{fig:align-gridlock}
\end{figure}

Low-mass galaxies (within halo of mass $<5\times10^{11}\, \rm M_\odot$) show some preferential alignment along the $x$-, $y$- and $z$- axis of the simulation box. 
In order to evaluate the effect of grid-locked galaxies on the galaxy--filament alignment signal, we have removed galaxies whose spin is comprised within less than 10$^\circ$ of any of the Cartesian planes of the box.
Fig.~\ref{fig:align-gridlock} shows that the alignment signal without grid-locked galaxies is comparable to the case where all galaxies are accounted for.
Low-mass galaxies have spin preferentially aligned with their filament, and massive galaxies have a spin perpendicular to the filament with a transition mass between $10^{10.25}$ and $10^{10.75}\,Ê\rm M_\odot$.
This behaviour is expected as filaments do not suffer from grid-locking; the effect of grid-locking on low-mass galaxies brings some extra noise to the alignment measurement.
Thus, the signal obtained for alignment of low-mass galaxies, while probably underestimated, is a robust trend.
The same is true for high-mass galaxies that do not suffer from spurious grid-locking.

\end{document}